# 3D Modular Microrobots: Micro-Origami Cubes with Integrated Si Chips Dive, Communicate, Flash Programs, and Form Collectives

Yeji Lee[1,2], Vineeth K. Bandari[1,2]*, John S. McCaskill[1,2,3]*, Pranathi Adluri[1,2], Daniil Karnaushenko[1,2], Dmitriy D. Karnaushenko[1,2] & Oliver G. Schmidt[1,2,3]*

[1]*Material Systems for Nanoelectronics, Chemnitz University of Technology, 09107 Chemnitz, Germany.*

[2]*Research Center for Materials, Architectures and Integration of Nanomembranes (MAIN), Chemnitz University of Technology, 09126 Chemnitz, Germany.*

[3]*European Centre for Living Technology (ECLT), Ca' Bottacin, Dorsoduro 3911, Venice, 30123, Italy.*

*Correspondence and requests for materials should be addressed to V.K.B. (email: vineeth-kumar.bandari@etit.tu-chemnitz.de), J.S.M. (email: john.mccaskill@main.tu-chemnitz.de) or O.G.S. (email: oliver.schmidt@main.tu-chemnitz.de).

## Abstract

Modular microrobotics can potentially address many information-intensive microtasks in medicine, manufacturing and the environment. However, surface area has limited the natural powering, communication, functional integration, and self-assembly of smart mass-fabricated modular robotic devices at small scales. We demonstrate the integrated self-folding and self-rolling of functionalized patterned interior and exterior membrane surfaces resulting in programmable, self-assembling, inter-communicating and self-locomoting micromodules (smart-lets ≤ 1 mm3) with interior chambers for on-board buoyancy control. The microbotic divers, with 360 ° solar harvesting rolls, function with sufficient ambient power for communication and programmed locomotion in water via electrolysis. The folding faces carry rigid microcomponents including silicon chiplets as microprocessors and micro-LEDs for communication. This remodels modular microrobotics closer to the surface-rich modular autonomy of biological cells and provides an economical platform for microscopic applications.

## 1. Introduction

The 3D folding of proteins and thin membranes into modular architectures supports a dense integration of mass-fabricated autotrophic and autonomous functions on accessible surfaces in cellular life (1). Functionally, this 3D deployment of self-folded, component-rich, active surfaces, which can selectively interact internally or with the environment, far exceeds the capabilities of monolithic chip technology, notwithstanding progress in the construction of on-board electronically controlled microrobots (2). Modular microbotics requires the integration of natural power harvesting and storage, complex electronic control, sensors, and actuators beyond current levels of on-board control, especially in soft robotics (3-7). Heterogeneous integration without self-folding (8,9) has enabled progress in smart dust (10,11) and autonomous chiplets in solution (12,13); easing but not overcoming (14) the planar layering of massively parallel lithography (15) which has constrained progress in microrobotics. In this work, we realize a novel mass-fabricable, autonomous and programmable modular microrobotics platform, extending the joint power of monolithic lithography and parallel heterogeneous integration via biomimetic self-folding fully into 3D. We demonstrate its use in the simultaneous integration of ambient power harvesting, module intercommunication, self-assembly and on-board electronic programming and control of microbotic locomotion, envisioned as smartlets in (16). Without this complete integration by self-folding, there is insufficient exterior surface area on autonomous self-powering microrobots at ≤1 mm for multi-faceted docking and peer-to-peer communication. Aquatic divers in 3D, autonomous unlike the famous Cartesian diver, known since Galileo's time, exemplify this breakthrough.

Self-folding of micro-origami (17-19) and Swiss-rolled (3,14) functional surfaces will be shown to convert structured membranes carrying bonded chiplets into penta-functionalized microbots with accessible interior functional structures. Micro-origami has been used to make passive free standing (20,21) and electronically active microstructures (19,22) as well as untethered microgrippers (23) and surface-tethered microrobots (24). Power has however limited the functionality of autonomous microdevices despite substantial developments in the microscale integration of radio frequency (25), optical (26) and ultrasound (27) radiation for energy harvesting and communication. Electronic actuators (28), especially for mobility, sensors (29) for environmental awareness, and transistor electronics for microcontrol on robots (2), compete for energy as well as space. The free-standing smartlets introduced here, with 360° solar harvesting rolls, function with sufficient ambient power for locomotion in water via electrolysis. The folding faces carry rigid microcomponents including silicon chiplets as microprocessors and micro-LEDs for communication.

## 2. Chiplet Integration in 3D-folded modular microbots

The compact access to energy is achieved in this work by self-folding and self-rolling thin film surfaces, which then demands that control electronics be integrated flexibly, too. While thin-film transistors (TFTs) are a natural choice, their miniaturisation still lags (30) far behind crystalline CMOS, consuming too much power for complex robotic control (31). Single crystal chiplet microcontrollers with microscopic dimensions have been produced (32), but for use on folded micro-origami robots there is still the major challenge of parallel reliable fine-pitched electrical connection to thin conducting traces which must robustly cross folds of their thin supporting surfaces. Here, we advance the flip-chip bonding of microscale Si-CMOS chiplets and other microelectronic devices (down to 100 μm scale) to such thin films, not only standard metallized polyimide (typically ≤5 μm thick) but also more complex multilayer films including hydrogels required for micro-origami (18,19), allowing such devices to be deployed as rigid islands on the flexible faces of the micro-origami cubes as shown in **Fig. 1**, independently of the rolled structures on the edges used for solar harvesting.

The integration of micro electronic sensors, actuators, optical communication elements alongside planar bonded chiplets on thin films using sequential micro-rolling (3,14) and micro-origami self-assembly (19) techniques results in a compact multifunctional smartlet, including an omnidirectional solar power source, without compromising functionality. The structure is reliably assembled in 3D, transforming it into a compact cube adorned with 8 tubes on its 8 edges, as illustrated in **Fig. 1a-c** (see [**SI Video 1**](#) & [**SI Video 2**](#)). **Fig. 1c** is a photo of the programmable multifunctional smartlet. Five functional components enabling autonomous robot functionality, visible at different orientations of the smartlet cube in Fig. 1d, have been integrated (two quantified in **Fig. 1**): (i) omnidirectional energy harvesting by micro-organic solar cells (μOSCs); (ii) orientation-sensing by micro-organic photodetectors (μOPDs); (iii) a custom programmable silicon CMOS chiplet (μChip) (12,13), controlling the smartlet on-board (iv) μLEDs in different colours enabling optical communication between smartlets; (v) bubble generating electrodes (BGEs) splitting water into gas bubbles ($H_2$, $O_2$) within the cube, reversibly bestowing controlled buoyancy upon the smartlet under the control of the μChip program. All the external faces of the smartlet cube are unoccupied, allowing them to be used for modular self-assembly control. The mass-fabrication results in cut tapes of many smartlets (**Fig. 1e**).

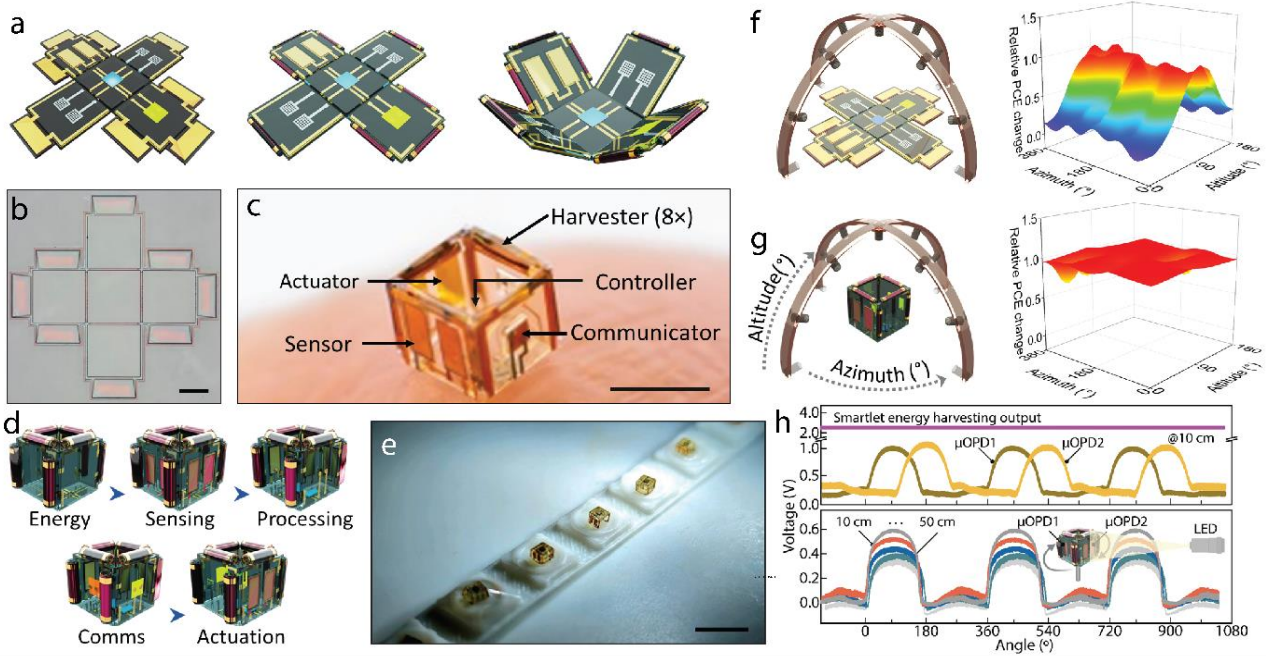

**Fig. 1 Functional integration and self-powering of programmable modular robotic cubes (smartlets). a.** Schematic illustration of the sequential rolling and folding self-assembly of the smartlet, integrating all functional components. **b.** Patterned carrier layer (polyimide) with faces of cube flanked on all 8 exterior edges by bevelled rectangles for the micro-organic solar cells (μOSCs) (Scale bar, 0.5 mm). **c.** Multi-functional programmable micro-origami cube, resting on a fingertip, integrated with an eightfold μOSCs energy harvester for power, micro-organic photodetectors (μOPDs) as sensors for use as orientation detectors and as communication receivers, a pulsed micro-LED (μLED) as sender for optical communication, bubble generating electrodes (BGEs) operating via a photoelectrochemical mechanism as actuators, and a silicon micro chiplet (μChip) as the on-board controller with 180nm resolution CMOS transistors for programming and control. **d.** CAD rendering of the assembled smartlet with the key successive functionalities in this work, shown in the relevant cube orientations. **e.** Linear tape containing smartlets fabricated photolithographically and self-assembled in parallel, apart from the pilot serial chiplet and LED bonding process. **f-g.** Schematic measurement setups for and surface plots of the relative change in power conversion efficiency (PCE) of a smartlet as a function of incident light orientation in 3D in both the pre-folded **(f)** and post-folded **(g)** state. Measurements employed an illumination intensity of 50 μW cm$^{-2}$. **h.** Top (purple): Orientation independence of solar harvesting with entire smartlet (8 deployed μOSCs) with schematic measurement (inset), contrasted with orientation sensitive orthogonally directed photodetectors μOPD1 and μOPD2 as the smartlet is continuously rotated around the symmetry axis of the smartlet. Below: The μOPD response as a function of angle is measured for three full rotations at five distances.

## 3. Compact rolled surfaces for omnidirectional ambient solar powering

Naturally occurring unconcentrated sunlight is limited to 1 mW/mm2 (1 sun) and should ideally be captured at all microrobot orientations. The P3HT: PCBM μOSCs used here are flexible, integrable, and theoretically capable of 33% power conversion efficiency (33). By integrating these materials onto pre-strained polymeric films, we created microtubes whose performance is independent of azimuthal light angle upon release (see **SI Fig.1**). A single rolled μOSC can generate a power conversion efficiency (PCE) up to 11.5 % at a short circuit current (Isc) of 50 μA and an open circuit voltage (Voc) of 0.65 V. Strategically positioning eight rolled μOSCs along the eight edges of the smartlet allows for omnidirectional energy harvesting (**Fig. 1c**) while keeping the smartlet faces available for microelectronic integration. By connecting the eight μOSCs in series, this design achieves a PCE of up to 1.5% without bypass diodes, with an output of Isc = 7 μA and Voc = 2.1 V (~17 μW). This energy is sufficient to power all functionalities of the smartlet simultaneously in this work. Moreover, it allows the smartlet to move freely, harvesting ambient solar energy regardless of orientation. The omnidirectional light-harvesting of smartlets is analysed in 3D (**Fig. 1f,g**), comparing the angle-dependent response of the unfolded structure's μOSCs with the isotropic response of the folded smartlet.

Awareness of orientation is important for smartlets for docking or navigation. The two μOPD photosensors mounted on one face of the cube, **Fig. 1h**, exhibit systematic voltage variation in response to changed orientation or distance of incident light (e.g. from other smartlets or the one-sun source). μOPDs on multiple faces (**SI Fig. 2**) can measure a wider range of orientations simultaneously, as shown in **Fig. 1h**. Additional detector faces readily available on docked modules, controlled orthogonal self-rotation, or additional reference beams e.g. from fixed smartlets, may in future be employed to sense full 3D orientation.

## 4. On board programming with peer-to-peer communication by light

The integration of CMOS chiplets into fold-up smartlets, delivers powerful on-board microcontrol, pioneering the elevation of soft folded 3D thin films to complex programmable microrobots. In **Fig. 2** we demonstrate this programmability in connection with optical communication between smartlets. Communication between smartlets is needed to enable them to coordinate actions, share critical information, or collectively tackle complex tasks. As opposed to single binary or analog signals, which can be realized physically in simpler interacting systems, peer-to-peer communication of highly specific binary encoded commands and whole programs, as demonstrated here, is unprecedented at this scale, opening the potential for sophisticated programmed collective dynamics in future. Optical communication, while attractive in high speed, large bandwidth, minimal interference, and low signal attenuation, and successfully miniaturized (26), has been hitherto limited in microrobotics by the power needed for optical transmission. Two programmable chiplets are employed here: the lablet microprocessor chiplet (12-13), designed to deliver complex temporal signal patterns to multiple electrodes for electrochemical and electrokinetic manipulation, μChip1 in **Fig. 2a,b**; and a commercial LED driver chiplet (34), useful for the higher currents of green LEDs and larger actuators, μChip2 in **Fig. 2c,d**. In both cases, serial programs are loaded bitwise and stored in shift registers, to control digitally the temporal sequence of driving voltages on several actuator pads as demonstrated in **Fig. 2b,d**.

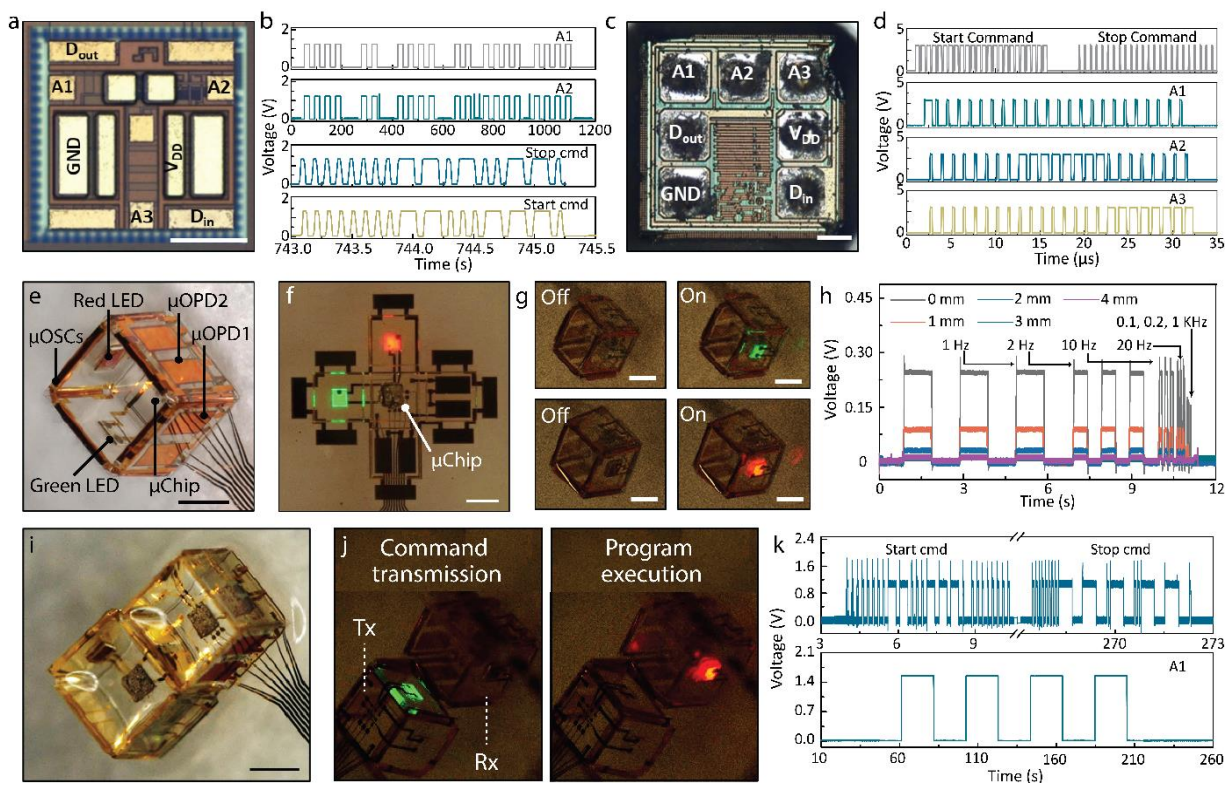

**Fig. 2 Smartlet on-board digitally logic and optical communication. a.** Microscope image of μChip1, a custom microcontroller CMOS chiplet, 140x140x50μm³ displaying the i/o interfaces used here: actuator outputs A1-3, program input $D_{in}$ and output $D_{out}$, ground GND, power supply $V_{DD}$ (Scale bar, 50 μm). **b.** Programming and output wave forms

of µChip1 after bonding to flexible substrate with Cu/Sn pillar bumps **c.** Microscopic image of µChip2, a 3-color (24-bit) LED driver chiplet WS2812B43, 300x300x100µm³ displaying all the i/o interfaces A1-3: actuator outputs, D-in/out: program input and output, GND: ground, $V_{DD}$: supply input (Scale bar, 50 um). **d.** Programming and output wave forms of the driver chiplet post-bonding. **e.** Microscope image of the smartlet, integrated with µOSCs, µOPDs (µOPD 1 positioned on the bottom face, µOPD 2 positioned on the right face), µChip and green and red µLEDs positioned on the left and top faces (Scale bar, 0.5 mm). **f-g.** Microscope image of the pre- **(f)** and post- **(g)** self-assembly smartlet displaying LEDs controlled by µChip: green (transmitter connected to $D_{out}$ of µChip2) and red (indicator connected to A1 of µChip1 or µChip2) (Scale bar, 0.5mm), see [SI Video 3](). **h.** Inter-smartlet communication: Operating bandwidth and distance characterization of the µOPD photosensors' response to a green light transmitting smartlet at 0 to 4 mm. **i-k.** Two optically communicating smartlets: **i.** Microscope bright image (Scale bar, 0.5 mm). **j.** Low light images of transmitting ($T_x$ green) and receiving ($R_x$ red) smartlets in 3 rows: Tx sending start command, Rx execution of its program, $T_x$ sending stop command, see [SI Video 4](). **k.** Top: Optical start and stop commands as specific pulse sequences generated by reception of programmed light directed at the µChip and Bottom: actuation signal (A3) initiated (activated) by the start command and stopped by the stop command.

Two pairs of µOPDs and two µLEDs are mounted on four faces of the smartlet in **Fig. 2e**. The green transmitter and red indicator LEDs, addressed by the µChips shown in **Fig. 2f,g**, can not only be used to visualize the internal actuation generated by the µChip program, but also to send data or commands from one smartlet to the next ([SI Video 3]()). The two photosensors on each face, connected in series, can generate a signal of 1-1.2 V useable as a pulsed digital input to the µChip data input channel, decoded as a command or program for the chiplet. In Fig. 2h, the green LED on a single smartlet is capable of transmitting data at rates from 1 to 1000 Hz, conveying digital information to a second smartlet through the aqueous media at separation distances under 4 mm without physical contact shown in **Fig. 2h**. This distance would in future be sufficient to support local communication during docking or in swarms of smartlets. **Fig. 2i-k** demonstrates the practical communication between smartlets of functional information such as start and stop commands. On the receiving smartlet (Rx), its programmed red-light pulses are started and stopped by commands encoded as green light from the sending smartlet (Tx) ([SI Video 4]()).

## 5. Intercommunicating on board control of microbotic aquatic divers

We address the on-board robotic motion control of smartlets in an ensemble in **Fig. 3**. Although external actuation is useful in some scenarios, independent agents require the more challenging control and powering autonomy of on-board actuation – beyond external magnetic fields (35) or light (2) controlling individual locomoting microrobots. Full power autonomy utilizes replenishable on-board resources, unlike metallic catalytic motors (14,36). The chosen actuator mechanism is one that only requires water and should be applicable in many natural environments, using µOSCs in series to create an electric potential sufficient to split water. To ensure efficient electrochemical actuation, nickel (Ni) and platinum (Pt) coatings are employed to catalyze the oxygen and hydrogen evolution reactions, respectively, as illustrated in **Fig. 3a**. Successful integration of such bubble generating electrodes (BGEs) into the smartlet is shown in **Fig. 3b-d**, in action together with µOSCs in **b** ([SI Video 5]()), and then in combination with the other functional components, before and after self-assembly, in **c** and **d** respectively.

Directed inter-smartlet local light programming, as depicted in **Fig. 3e**, represents a major advance in actuation control for modular robotics at this scale. Smartlet programs on the attached lablet µChip are 58 bits in length and encode a state machine which has 4 operating states (idle, programming, running, sending) and is sensitive to external 8-bit commands while idle or running. **Fig. 3f,g** shows the ensuing locomotion, with one smartlet (S2) optically transmitting a command, "start your locomotion program", to a second smartlet (S3) in the presence of a third (S1). This program initiation triggers a series of actions in the programmed smartlet S3. The smartlet begins generating bubbles inside and once a critical gas volume accumulates, the smartlet levitates, free to rotate and laterally translate, propelling upwards. After reaching the surface, the on-board program halts the bubble generation, and remaining bubbles start to dissolve (37), reducing smartlet buoyancy, causing it thereafter to return to the vessel floor (SI Video 6). This controlled movement, with timed surfacing cycles, demonstrates the potential for sophisticated motion control enabled by the programmable µChip.

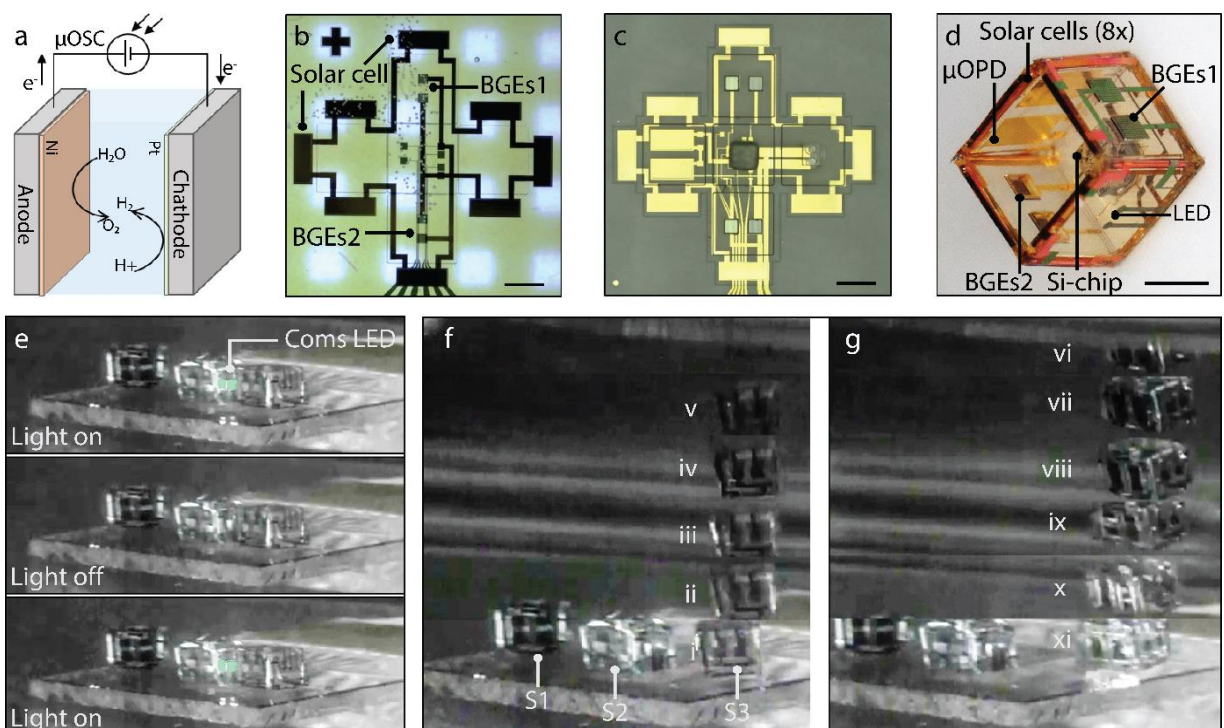

**Fig. 3 Programmed locomotion in water via on-board digital logic. a.** Schematic of the working mechanism of electrochemical water splitting into $H_2$ and $O_2$ gas bubbles at Pt and Ni coated electrodes. **b.** Microscope image showing bubble generation at two integrated electrochemical bubble generating electrode pairs (BGEs, the Pt and Ni electrodes) powered by µOSC solar cells under standard solar illumination (1 sun) in an unfolded smartlet (Scale bar 0.5 mm). For full time sequence see SI Video 5. **c-d.** Microscope close-up of the smartlet, pre- (**c**) and post- (**d**) self-assembly, showing organic solar cells, photodiodes on left side, a microchip at the centre, LEDs on the right side, and electrochemical BGEs on both the top and bottom faces of the cube (Scale bar 0.5 mm). **e.** Snapshots demonstrating three frames (on-off-on) taken from active smartlet communication via green µLED on the transmitting smartlet ($T_x$) to the receiving smartlet ($R_x$) µOPD sensor. **f-g.** Real-time image overlay sequence illustrating how electrochemical buoyancy engines via BGEs, switched on and off by the on-board program, cause the smartlet S3 to move first up (i-v) (**f**) and then down (vi-xi) (**g**). In this case the tethered smartlet S2 sends the start command via its green µLED (as shown in **e**) to initiate the autonomous programmed locomotion sequence up and down repeatedly with intermissions on smartlet S3. For full time sequence, see SI Video 6.

**Fig. 4** demonstrates both the selective in ensemble and local communication needed for modular robotics. Firstly, different digital IDs and communication frequencies of variant CMOS chiplets also enable the selective global addressing and hence motion control of individual smartlets within a collective, as demonstrated in **Fig. 4a,b,** where smartlets S2 and S3 are untethered but respond to programs and commands issued at different data transmission rates. Only smartlet S3 is programmed in response to the global 200 Hz light-encoded command: it ascends to the surface where it stays for some time. While the large excess buoyancy of S3 slowly depletes via bubble solvation, smartlet S2 is programmed to navigate repetitively to the surface and back to the floor twice, with a smaller excess buoyancy, before receiving a 50Hz encoded stop command and returning permanently to the floor. Overall, local (inter-smartlet) and global light commands, both demonstrated for smartlets, provide programmable functions for coordinating and controlling collective movement (38). It is expeditious to separate this initial breakthrough from extensive follow-on work needed to exploit the rich collective behavior in larger numbers of smartlets.

## 6. Self-assembly of smartlet modules

Active modular robots can assemble into larger structures by controlled docking at cm scales (4,39). While the self-assembly of passive modules by hydrophobic-hydrophilic patterning (40,41) in the form of mm-scale disks (42), magnetic bar code patterning (43), or specific DNA hybridization on gel cubes (44), with the option of environmental switching of docking interactions, has been achieved with recourse to mechanical shaking at scales where inertia dominates Brownian motion (≳1 μm), active self-locomotion assisted assembly for microbots at the 1 mm-scale or below has not. The dynamic interplay of long range locally programmed smartlet self-locomotion and short range self-assembly is demonstrated in the sequence of still frames in **Fig. 4c,d,** with full video documentation provided in the [SI video 7](). Smartlets navigate to the liquid-air interface autonomously, under the control of their on-board electronic program, where now buoyant they are attracted laterally to one another by the surface tension force acting on their displacement menisci. They approach and dock. Depending on the programmed individual or synchronized diving of smartlets (through bubble generators being turned off at different or the same times) either specific undocking (    ) or return of the assembly to the beaker floor occurred. This is a first demonstration at the 1 mm scale of the electronic self-locomotion enhanced self-assembly of modular robots.

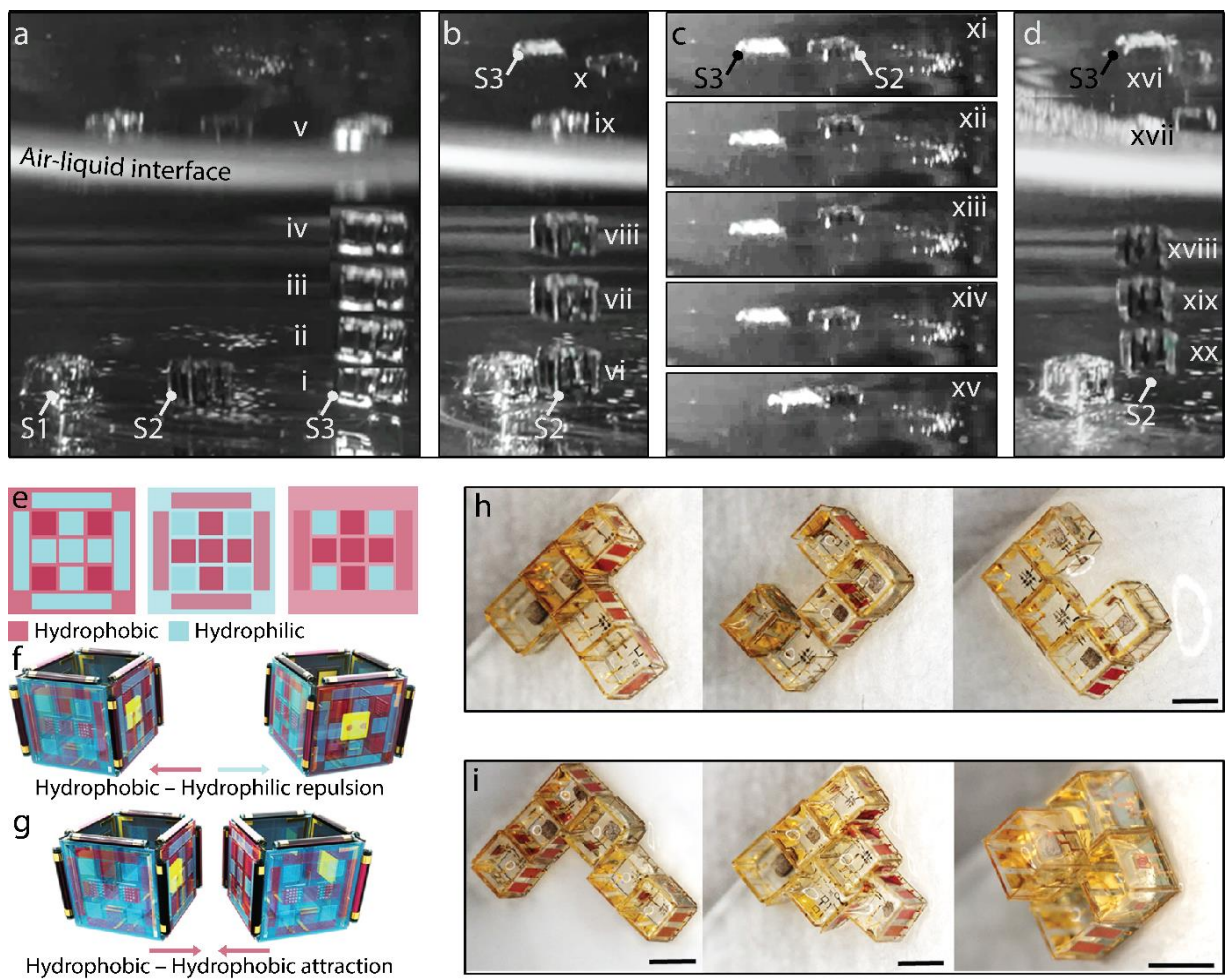

**Fig. 4 Smartlet collective motion and guided self-assembly. a-b.** Real-time image overlay sequence depicting frequency-selective, individually instructed smartlets S3 and S2, both untethered, commencing locomotion programs selectively in response to command data: **(a)** S3 sent at 200 Hz (i-v) and **(b)** S2 at 50 Hz (vi-x). **c-d**. Autonomous active docking and undocking of two smartlets assisted by surface tension. In **b**: The second smartlet S2 locomotes from the glass beaker bottom to the air-water interface with excess buoyancy under the control of its on-board program (overlaid frames vi-x), near to where the first smartlet S3 has already arrived. Smartlet S1 provides a stationary point of reference. In **c**: After initial kinetic repulsion (xi-xii), and then capillary attraction (xiii-xiv) the two smartlets dock together (xv) with the assistance of surface tension. In **d**: Under the active control of its on-board program, smartlet S2 ceases bubble formation, reducing its buoyancy to overcome the surface tension forces holding it to smartlet S3 and returns to the beaker bottom (xvi-xx). For full time sequence see [SI Video 7](). **e**. 3D CAD drawings of the hydrophobic-hydrophilic patterning of the outer surfaces of the smartlet cubes. **f-g**. Repulsion of non-matching and attraction of matching patterned smartlet faces in water. **h-i.** Partially guided self-assembly of smartlets: **h.** into the letters "T," U" and "C". Linear self-assembled primitives of 2-3 smartlets are brought together with laterally placed smartlets which dock by self-assembly to complete the structures. **i.** Self-assembly of other structures, including half-registered contacts, enabled by the registration controlling rectangles (Scale bar 1 mm).

But can the self-assembly of such multi-functional robotic smartlets be also made face specific? Through specific hydrophobic-hydrophilic 2D-barcode sub-patterning of the outer faces of the smartlet, specific self-assembly, based on matching surface patterns and self-locomotion has for the first time been achieved at the 1 mm scale for fully functional electronic robot modules, see **Fig. 4e-i**. The exclusive use of inner faces of the smartlet cubes for functionalization, with light harvesting relegated to edge rolls, leaves the outer faces available to control docking as shown schematically in **Fig. 4e-g**. However, these surfaces, which face down towards the sacrificial layer in micro-origami fabrication, are not straightforward to pattern hydrophobically. We employed e-beam deposited $MoO_3$ structured by lift-off as a water-soluble sacrificial layer for a special precursor layer, which only later becomes hydrophobic after heat treatment. These patterns are masked via planar lithography, maintaining the mass-producible nature of the assembled smartlet robotic modules (see Methods for details).

Smartlets self-assemble to specific modular structures, with docking of matching cube faces, when placed in the vicinity of one another. The autonomous long-range transport required to bring appropriate aquatic divers to such vicinity, in the absence of macroscopic agitation (41-44), has been demonstrated above (**Fig. 4c**). Examples of assembled multi-module structures are shown in **Fig. 4h,i**. They are held together by the standard hydrophobic/hydrophilic patterned forces for self-assembly in water (41). Smartlets first self-assembled to primitive shapes, like short linear chains or square arrays. To emulate the long-range transport recycling of additional smartlets in global self-assembly, considering the slow diving cycles of the current smartlets, we simply added them manually into the vicinity, where they self-assembled onto the above assemblies. Both completely face-registered (**Fig. 4h**) and half-offset (**Fig. 4i**) self-assembled structures were achieved through the design of the surface tension controlling patterns to include peripheral rectangular registration guiding elements (see **Fig. 4e**).

## 7. Discussion

This integration of autonomous natural powering and on-board electronic control of locomotion and inter-module communication takes peer-to-peer microrobotic control beyond simpler physicochemical mechanisms (45). Although our robot demonstrations are simple, the pathway is now wide open to engineer, with incremental design, improvements to functionality and further miniaturization. A chiplet tuned for modular robotics, with enhanced functionalities and memory of current CMOS, will soon extend the binary sensor feedback control in lablets to the higher temporal and signal strength resolution required for complex robotic action. The bonding of chiplets onto the in-folding surfaces allows a large increase in functional surface area inside microrobots, freeing exterior surfaces for multi-module docking interactions – both by self-assembly and active locomotion. Internal buoyancy-control has already proven to be an efficient mechanism of powering underwater gliding locomotion in ocean monitoring under-water drone vehicles (46). The on-board lablet programs already support multi-actuator control (13,47), making independent control of the two on-board bubble generators straightforward, permitting left-right locomotion control through tilt, and hence 3D controlled movement of smartlets equipped with gliding flaps.

## 8. Conclusions: scalability of the technology

As a modular, mass producible entity (flip-chip-bonding of chiplets can also be fully automated), permitting rapid rounds of development and parallel testing, we suggest that smartlets can now be down-scaled systematically following Wright's Law (48,49). The conditions are no fundamental physical barriers to complex optimization, and the need for, fabrication and testing of exponentially growing numbers of modules with time. Currently, scaling starts from 2" wafers with 64 1-mm smartlets, with the next pipelined step of 500μm smartlets on 4" wafers increasing this to 1024 per wafer. The CMOS chiplet fab yields over 100,000 per 8" wafer. In collective functionality, demonstrated here with both individually controlled locomotion and self-assembly as well as advanced inter-communication of modules, lies the full potential of smartlets. Consideration of communication, locomotion, assembly and energy scalings reveal no insurmountable physical barriers to miniaturization from 1mm to 20 μm, the scale of cells. For example, while the available solar power decreases with area, so do the power needs. The chiplet controllers employed will fit easily on 250 μm scale smartlets, resting power < 2 nW, and can be scaled down to 25-50 μm and sub nW power using current 12-22nm resolution CMOS. Ten times harvesting efficiency, gainable by an integration of transistor diodes with serial solar cells and by material optimization, would then downscale to deliver 28 nW solar power for a 20 μm smartlet. Size and power needs of actuators scale down at least proportional to area, and for local communication, μLEDs are already available at 50 μm. It is the modular approach which opens the path to million-fold fabrication and testing, that enables regular halving in size of modules. The integration of active information with smart particles has broad applications from minimally invasive medicine to information-optimized recycling of docked products. This contribution sets the stage for autonomous modular electronic microrobotics, paving the way to its many applications (16).

## 9. Methods

### 9.1 Fabrication of light sensitive structures: μOSC and μOPD

The fabrication of μOSC (micro-organic solar cells) and μOPD (micro-organic photodetectors) on a polymeric platform is conducted through a streamlined process where both devices are simultaneously fabricated, sharing several essential elements (see **SI Figs 3-5**). Analysis regarding morphology, energy band gap alignment diagram and material characteristics are provided in **SI Figs. 6-9.**

Bottom contact pattern: The process begins with maskless photolithography using AZ 5214 E photoresist (MicroChemicals) to define the electrode structure. Spin-coating the photoresist at 4500 revolutions per minute (RPM) for 45 seconds to attain 1 μm thickness and soft baking at 90 °C for 4 minutes, followed by UV exposure (365 nm, 15 mW/cm2) for 5 seconds. After exposure, samples are post baked at 120 °C for 2 minutes, undergo full UV exposure for 30 seconds and developed in AZ 726 MIF developer (MicroChemicals) for 45 seconds. 10 nm and 50 nm thickness of chromium (Cr) and gold (Au) films respectively are sequentially deposited onto the patterned substrate at a controlled rate of 0.5 Å/s using an electron-beam evaporator (Creative Vakuumbeschichtung GmbH). A lift-off process is employed thereafter to remove non-exposed photoresist and residual metal layers, yielding a cleanly patterned bottom contact layer crucial for the integration and operation of μOSCs and μOPDs.

ITO cathode: Preparation for subsequent processing includes negative lithography using AZ 5214 E photoresist onto the electrode contact layer. A 100 nm layer of indium tin oxide (ITO, EvoChem) is then deposited at 0.2 Å/s using a magnetron sputtering machine (Moorfield). Following ITO deposition, a lift-off process removes photoresist and any residual metal layers using a mixture of acetone and isopropyl alcohol (Sigma Aldrich). The samples are annealed at 200 °C for 5 hours to stabilize the ITO layer and enhance its conductivity, thereby improving the efficiency of μOSCs and μOPDs.

Electron transport layer: Zinc oxide (ZnO) serves as the electron transport layer (ETL), deposited by thermal atomic layer deposition (S100 from Savanna). The process employs dimethyl zinc (DMZ) as the precursor and water (H2O) as the oxidizing agent at a substrate temperature of 200 °C. Deposition is carried out at a rate of 30 Å/cycle. Positive lithography is subsequently used to pattern the ZnO layer, which is then etched with an 85% concentration in volume of phosphoric acid (H3PO4, MicroChemicals). A post-deposition annealing at 100 °C for 20 minutes enhances the electrical and structural properties of the ZnO layer.

Photoactive layer, hole transport, and anode layer: The photoactive material comprised of a blend (in 1:1 ratio) of poly(3-hexylthiophene) and [6,6]-phenyl C61-butyric acid methylester (P3HT:PC61BM, Sigma Aldrich) is dissolved in 1,2-dichlorobenzene (Sigma Aldrich) at a 20 mg/ml concentration. The solution is spin-coated at 800 RPM for 90 seconds through a 0.2 μm PTFE filter, and baked at 140 °C for 10 minutes to form a 200 nm thick film. Methanol immersion for 10 minutes post-spin improves film morphology. The hole transport layer (HTL) made from poly(3,4-ethylenedioxythiophene) polystyrene sulfonate (PEDOT: PSS, CLEVIOS™ F HC Solar) is processed by spin-coating a filtered solution at 5000 RPM for 60 seconds, achieving a thickness of about 50 nm. Annealing at 120 °C for 10 minutes enhances electrical conductivity and film uniformity. All procedures are conducted in a nitrogen-filled glove box to maintain material integrity. The top electrode is fabricated by depositing a 120 nm layer of Au at a rate of 0.4 Å/s using an electron-beam evaporator (Creavac), followed by patterning through positive lithography. Au layer etching with a commercial Au-etching ACL2 solution (MicroChemicals) and O2 plasma etching (TEPLA) at 400 W for 5 minutes defines the device architecture. A lift-off process removes non-exposed photoresist.

Passivation: To protect the devices from the highly basic rolling solution, a protective layer of SU8-2 photoresist (Micro Resist Technology) approximately 1 μm thick is applied. Spin-coating the photoresist filtered with a 0.2 μm PFTE filter at 8000 RPM for 60 seconds ensures uniformity. Soft baking at 60 °C for 10 minutes solidifies the photoresist, followed by UV exposure (365 nm, 15 mW/cm2) for 65 seconds through a photomask. Post-exposure baking at 60 °C for 5 minutes further cures and stabilizes the photoresist pattern. Development in mr-Dev 600 (Micro Resist Technology) for 60 seconds removes unexposed photoresist areas, revealing underlying material while preserving the exposed, cross-linked photoresist as a passivation layer.

## 9.2   Fabrication of light sensitive structures: μOSC and μOPD

Layer Thickness and Surface Analysis: The thickness of individual layers were accurately measured using a surface profiler (Veeco Dektak 8). Morphological and surface feature assessments were conducted using optical microscopy (Olympus BX5) and scanning electron microscopy (SEM) with a GAIA3 TESCAN operating at an acceleration voltage of 5 kV Ga+.

Electrical and Photovoltaic Characterization: The current-voltage (I-V) characteristics of the devices were evaluated under standard intensity of 100 mW/cm² using a solar simulator (LSE341, LOT QuantumDesign GmbH), equipped with 150-600 W arc lamps. The simulated illumination intensity was calibrated using a commercial optometer (ILT2400, International Light Technologies). I-V curves were analysed using a source meter (Keithley 2636A) across a voltage range of -1 to 1 V for single cells and extended to 3 and 5 V for serially connected cells. Key parameters including short-circuit current ($I_{sc}$), open-circuit voltage ($V_{oc}$), maximum power ($P_{max}$), fill factor (FF) and power conversion efficiency (PCE) are derived from the I-V curve (see **SI Note 1**).

Reliability and stability characterization: Planar and tubular μOSCs were measured for three different devices in order to confirm the stable fabrication protocols and performance reliability, as shown in **SI Figs. 10-11**. The operational stability of both planar and tubular μOSCs was evaluated over extended periods under ambient conditions to ascertain their long-term reliability and sustainability. Throughout a 31-day testing period, tubular μOSC exhibited excellent stability, retaining up to 60% of their initial PCE and maintaining nearly constant $V_{oc}$, while being directly exposed to atmospheric conditions. In addition, planar μOSCs demonstrated retention of 90% under the same conditions (see **SI Figs. 12-13**).

Optical Characterization: Angle-dependent performance at varying light incidence angles was assessed (see **SI Fig. 14**) using an angle-changeable microscope (Keyence VH-Z 100R). The performance of the planar μOSC exhibits systematic variations in response to changing incident angles, as depicted in **SI Fig. 15**. When the incident light is perpendicular to the device surface, maximizing light absorption, the μOSC demonstrates peak performance. In contrast, tubular μOSCs respond differently to varying incident angles due to their unique 3D architecture, which facilitates efficient light trapping and absorption from multiple directions. Light illuminating the sides of the tubes can still be effectively absorbed and converted into electrical energy, as shown in **SI Fig. 16**. This angular versatility inherently enhances the overall efficiency of μOSCs across a wide range of incident angles.

Environmental Conditions: All measurements were conducted under ambient conditions to ensure the data reflects the typical operational environments for these devices.

### 9.3   Fabrication of self-assembling polymeric platform

The self-assembling polymeric platform provides the foundational structure essential for creating tubular μOSCs (micro-organic solar cells) and 3D micro-origami cubes. The process begins by patterning the self-assembling polymer stack as a planar structure, followed by the initiation of rolling and folding mechanisms by the selective release of sacrificial layers.

Patterning the self-assembling polymer stack, procedure adapted from (57): Initially, a glass substrate is treated with oxygen plasma (TEPLA) at 400 W for 2 minutes to enhance adhesion for the subsequent polymer layers. A lanthanum-acrylic acid-based organometallic photo-patternable material is then spin-coated at 3000 RPM for 90 seconds, soft-baked at 40 °C for 10 minutes, followed by UV exposure using a MA6 Mask Aligner (Süss, 365 nm, 15 mW/cm2) for 60 seconds, developing in DI water for 10 seconds and hard baking at 220 °C for 30 minutes, creating a sacrificial layer (SL) 300 nm thick. A photo-patternable hydrogel layer (HGL), serving as the hinge for the origami cube and a rolling platform for tubular μOSCs, is then spin-coated at 6000 RPM for 90 seconds, soft-baked at 40 °C for 10 minutes, followed by UV exposure using a MA6 Mask Aligner (Süss, 365 nm, 15 mW/cm2) for 90 seconds, development in diethylene glycol monoethyl ether (Sigma Aldrich) for 20 seconds and hard bake at 220 °C for 30 minutes, to achieve a thickness of 800 nm. A photo-patternable polyimide rigid layer is subsequently applied by spin-coating at 5000 RPM for 90 seconds, soft-baked at 50 °C for 10 minutes, followed by UV exposure using a MA6 Mask Aligner (Süss, 365 nm, 15 mW/cm2) for 70 seconds, developed in a mixture of 1 wt% Ethanol:2 wt% Diethylene glycol monoethyl ether: 4 wt% 1-Ethyl-2-pyrrolidinon (Sigma Aldrich) for 60 seconds and hard baked at 220 °C for 30 minutes, to reach a thickness of 500 nm. A final, thick layer of 10 μm SU8-25 (Micro Resist Technology) is spin-coated at 3000 RPM for 60 seconds, soft-baked at 95 °C for 4 minutes, followed by UV exposure for 80 seconds, post-baked at 95 °C for 2 minutes and developed in mr-Dev 600 for 60 seconds to be patterned into a square shape. This square structure forms the base of the cube and provides robust support during the folding and manipulation processes (see **SI Fig. 17**). Following the formation of the polymeric platform, μOSCs are fabricated as described earlier in the method section and as shown in **SI Fig. 18**.

Micro-origami self-assembly process: The self-assembly into both tubes and cube is initiated by preparing a 0.1 M solution of sodium diethylenetriamine pentaacetic acid (DTPA, Sigma Aldrich). The pH of the solution is carefully adjusted from 6 to 9 using sodium hydroxide (NaOH, Sigma Aldrich), a crucial step for controlling the etching rate and selectivity during the removal of the sacrificial layer. This adjustment influences the precision of the rolling and folding processes. Once self-assembly is complete, the self-rolled and folded device is immersed in deionized (DI) water for 20 minutes to eliminate any residual rolling solution. Gentle agitation during the washing ensures the cleanliness and structural integrity of the device, preparing it for further processing and evaluation. The successful integration of Swiss-roll μOSCs and self-assembly devices are shown in **SI Fig. 19**.

### 9.4   Analysis of smartlet integrated μOSCs and μOPDs

Electrical and photovoltaic characterization: The smartlet integrated µOSCs connected in series are analysed by recording I-V curves. Initially, the devices are evaluated before rolling and folding, with measurements taken across three different devices, as shown in **SI Fig. 20**. These devices demonstrated maximum efficiency when the incident light angle is perpendicular to the cell. However, when the µOSCs are integrated into the smartlet, only the tubular µOSCs positioned at the top can efficiently absorb light leading to a Voc of 2.1 V to 3.1 V instead of 5.2V, as illustrated in **SI Fig. 21**.

Omnidirectional photovoltaic analysis: To characterize the omnidirectional light-harvesting capability of the fully assembled cube, we designed a 3D printer (SLA 3D printer from Formlabs) hollow hemisphere and placed 16 NeoPixel LED strips along the azimuthal axis, resulting in an angular distribution of one LED for every 22.5 °. Each of the 16 LED strips consists of approximately 21 LEDs distributed along the altitude axis, resulting in an angular distribution of one LED for every 17.5 °. All LEDs were calibrated to output 50 µW/cm² with the projection aimed at the centre of the hemisphere. This setup is shown **in SI Fig. 22** and has been used in the experiments corresponding to Fig. 1f,g and SI Figs. 23-25. This configuration ensures uniform light distribution from multiple directions, where the LEDs were programmed and controlled to operate at arbitrary positions (see **SI Note 2**). To assess the individual solar cell performance depending on their position within the smartlet, each cell is separately contacted (see **SI Fig. 23**). **SI Figs. 24-25** demonstrate that when the incident light is directly or closely aligned with a particular µOSC, it exhibits dominant energy harvesting compared to µOSCs positioned further away from the light source. This positional dependence highlights the importance of spatial arrangement and orientation optimization for maximizing energy absorption and overall device performance. [SI Video 8](#) shows the real-time of positional and orientational measurement of smartlet.

## 9.5   Preparation for SLID bonding

Although conductive epoxy-based transfer bonding methods are well-demonstrated (46,47), we present, for the first time, an alternative approach by extending the solid-liquid interdiffusion (SLID) bonding technique to ultra-thin polymeric films (thickness less than 5 µm). The Cu and Sn based SLID bonding method significantly reduces resistance compared to traditional epoxy-based transfer bonding techniques, offering a substantial improvement in electrical performance. For the preparation of the SLID bonding, which is crucial for integrating µLEDs and µChips, copper (Cu) and tin (Sn) are selected as conductive and solder materials due to their established properties in bonding applications. Cu and Sn layers are deposited using a standard electroplating technique, achieving thicknesses of 10 µm and 5 µm, respectively. After deposition, a reflow process is carried out under vacuum conditions to form well-defined solder bumps, thereby enhancing bonding efficiency (see **SI Fig. 26**). The alignment and bonding of flipped chips are performed using a commercial bonding machine (FinePlacer Pico), adhering to precise pressure and temperature profiles to ensure reliable and precise connections between components. This is critical for the functionality and performance of the integrated devices. The adaptation of advanced SLID bonding technology to polymeric thin film substrates facilitates the reliable bonding of both custom CMOS chiplets—nW-powered 140x140x35 µm microcontrollers with 12 or more pins—and simpler chiplets, including micro-LED chips (see **SI Figs. 27-28**). The sequential bonding process of the µChip and µLEDs are described in **SI Fig. 29** and [SI Video 9](#).

## 9.6   Lablet µChips

The description and programming of the custom lablet µChip1 and commercial LED driver µChip2 are described in **SI Note 3-6**. The short summary of lablets here gives readers the necessary orientation on their nature and use. Lablets were thinned and singulated from combinatorial variant array chips (each containing ca. 1000 lablets) designed at the Ruhr Universität Bochum and fabricated at wafer scale at TSMC in 180nm CMOS32. The 58-bit program for the lablets is serially loaded in a shift register and species details and timing for four-phases of processing, as temporal patterns of tristate or active signals on specific subsets of the 3 dedicated and 3 facultative actuator electrodes, as well as conditional switching between the phases in dependence on digital signals derived from the two electrochemical sensors. Other features of the programmable finite state machine, including the ability to record sensor data and to replicate itself between lablets are not used in the current work.

### 9.7 Operation of μLEDs

Detailed operational description of the green and red μLED is discussed in **SI Note 7**, including emission characteristics, voltage requirement and spectral properties. Following the SLID bonding process, successful bonding is confirmed through the illumination of μLEDs, demonstrating their functionality even with different colour combinations, as shown in **SI Fig. 31**.

### 9.8 Fabrication of BGEs

The complete fabrication of a smartlet begins with the creation of the polymeric origami platform, followed by BGEs, μOSC and μOPD layer deposition and patterning, SLID bonding of μLEDs and a μChip and finally SU-8 photoresist passivation, as illustrated in **SI Figs. 32-33**. Oxygen Evolution Reaction: To facilitate the oxygen evolution reaction, a thin layer of nickel (Ni), recognized for its photocatalytic properties, is deposited. The Ni layer is precisely patterned using negative lithography and deposited through magnetron sputtering at a controlled rate of 0.5 Å/s to achieve a thickness of 15 nm. Following deposition, a lift-off process using acetone and isopropanol is conducted to remove excess material and accurately define the desired pattern. Hydrogen Evolution Reaction: For the hydrogen evolution reaction, platinum (Pt) is utilized as the primary catalyst. Similar to the Ni deposition, negative lithography is employed to ensure the precise placement of the Pt layer. Prior to depositing Pt, a titanium (Ti) layer is applied using an electron beam evaporator (Creavac) to serve as an adhesive layer. The deposition rates for Ti and Pt are 0.2 Å/s, resulting in thicknesses of 5 nm and 10 nm, respectively. A lift-off process in acetone and isopropanol follows to remove any residual metal, thereby defining the clean and precise pattern required for effective catalytic activity.

### 9.9 Fabrication of hydrophobic layer for patterned smartlet docking control

This procedure is only employed for the customization of smartlet cube exterior faces for specific docking, as shown in **Fig. 4**. The preparation of the hydrophobic layer begins with AZ5214E negative lithography (details described above) on the SL to pattern areas corresponding to the faces of the smartlet, followed by a 200 nm thick amorphous MoO3 grown by electron beam deposition (Edwards) at a rate of 0.2 Å/s (58). Then a lift-off process was performed in acetone and isopropanol. This pattern served as a water-soluble sacrificial layer for patterning the hydrophobic material (ETC-PRO, from EVOCHEM), which was deposited to a thickness of 100 nm using electron beam evaporation (Edwards). Due to the good solubility of MoO3 in H2O, the hydrophobic materials were lifted off in DI water. As a result, the hydrophobic material that was not deposited onto the MoO3 layer remained on the substrate, forming a well-defined pattern. At the end of the fabrication sequence, a curing step was performed by heating the samples up to 170 °C for 2 hours to ensure the activation of the hydrophobic material. The schematic illustration of the patterning and folding process is shown in **SI Fig. 34**.

### 9.10 Experimental setup for controlled locomotion

The measurement setup to operate the fully equipped smartlets is illustrated in **SI Fig. 35**. It consists of a solar simulator to provide 1Sun calibrated light source and an LED light source at the top of the beaker as a global programming source. Smartlets are immersed into a 60 ml beaker filled with water and 0.075% methylene-blue to accelerate the dissolution of bubbles at the water-air interface. A logic analyser for programming the chiplets (digital discovery logic analyser) and two cameras (cam1: Basler acA1300-200uc; cam2: Keyence VHX digital camera) at different angles to capture the locomotion process from multiple perspectives are also part of the set-up.

**Contributions:** Conception: V.K.B. and O.G.S. led the project with conceptual input from J.S.M., Y.L. and D.K. Device fabrication: Y.L. and V.K.B. performed all steps from sample preparation to device measurements. Data analysis: Y.L., V.K.B., and O.G.S. Writing: Y.L., V.K.B., J.S.M. and O.G.S. with input from all authors. Custom chiplets: J.S.M. and O.G.S. conceived the low-power integrating role of, and J.S.M. provided and with V.K.B. supervised the singulation, programming and use of the lablet microchips by P.A. and Y.L. The CMOS lablets were created at wafer scale in the EU project #318671, conceived and coordinated by J.S.M. Polymeric layer stacks: D.D.K. and D.K. prepared the materials.

**Acknowledgements:** J.S.M. wishes to acknowledge especially the contribution of T. Maeke, D. Funke, P. Mayr, and J. Oehm in codesigning the electronics of the lablet μChips used in this work. We thank C. N. Saggau for absorption measurements and C. Schmidt, A. M. Placht, A. Dumler & P. Plocica for technical support and lab maintenance. J. Müller, press office TU Chemnitz, took the photo used in Fig. 1c, and P. Radha produced the animation in SI Video 1. The authors discussed the smartlet concept at TU Chemnitz with proposed partners in the Microelectronic Morphogenesis (MIMO) initiative. O.G.S. acknowledges financial support by the European Research Council (ERC) under the European Union's Horizon 2020 research and innovation program (grant agreement No. 835268). O.G.S. and D.K. acknowledge support by the German Research Foundation DFG (SCHM 1298/32-1, KA5051/3-1). This work was co-financed by tax revenue on the basis of the budget approved by the members of the Saxony State Parliament.

# Supplementary Information (SI):

**SI Figures 1-35**, & **SI Notes 1-7**.

**SI Video 1**. fabrication sequence and micro-origami enabled 2D to 3D transformation of the smartlet
( **Link:** https://tuc.cloud/index.php/s/tsSjGfRQFqXjr6n )

**SI Video 2**. Micro-origami enabled 2D to 3D transformation of the smartlet
( **Link:** https://tuc.cloud/index.php/s/tP2oaw9gr6bzSZn )

**SI Video 3**. Smartlet transmitting start and stop command using red LED
( **Link:** https://tuc.cloud/index.php/s/TrezfRyxdg3RGgY )

**SI Video 4**. Tx-smartlet to Rx-smartlet communication
( **Link:** https://tuc.cloud/index.php/s/o9NWBs7cyF7XPHz )

**SI Video 5**. Serial connected solar cells power water splitting at 1 sun
( **Link:** https://tuc.cloud/index.php/s/ceF4263TS7AXwX7 )

**SI Video 6**. Tx-smartlet to Rx-smartlet optical communication for programmed locomotion
( **Link:** https://tuc.cloud/index.php/s/xK6QmyH2xrGggNC )

**SI Video 7**. Frequency selective global light programming of individual smartlet
( **Link:** https://tuc.cloud/index.php/s/rxikTWgezW52kxs )

**SI Video 8**. Positional & orientational perception of smartlet
( **Link:** https://tuc.cloud/index.php/s/Gk4rBWam2qdPbgk )

**SI Video 9**. Alignment process for chiplet bonding onto Cu-Sn pillar bumps
( **Link:** https://tuc.cloud/index.php/s/fR646sdYxQ5Gp72 )

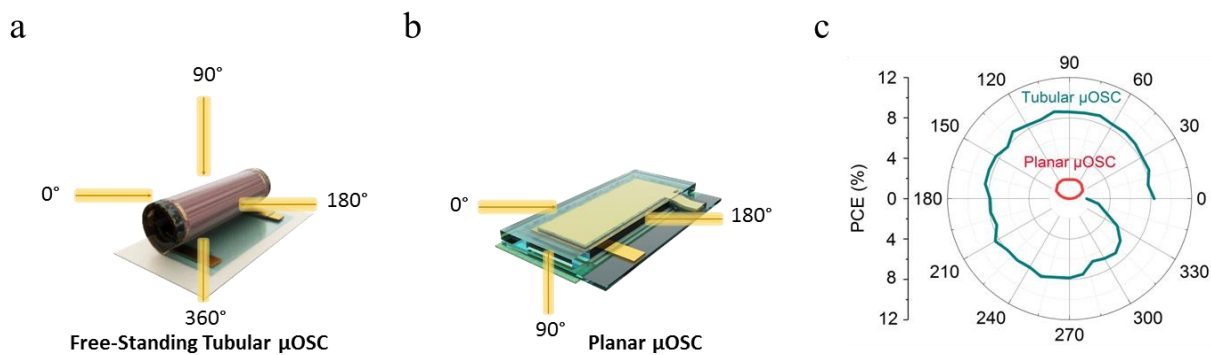

**SI Figure 1 | Incident angle dependent measurement of tubular and planar µOSC. a-b,** Schematic illustration of incident angle dependent measurement around single axis for **(a)** free-standing tubular µOSC and **(b)** planar µOSC. **c,** PCE retention of tubular and planar µOSC as a function of incident angles ranging from 0 ° to 360 °. All measurements are performed under the illumination of 100 mW cm$^{-2}$.

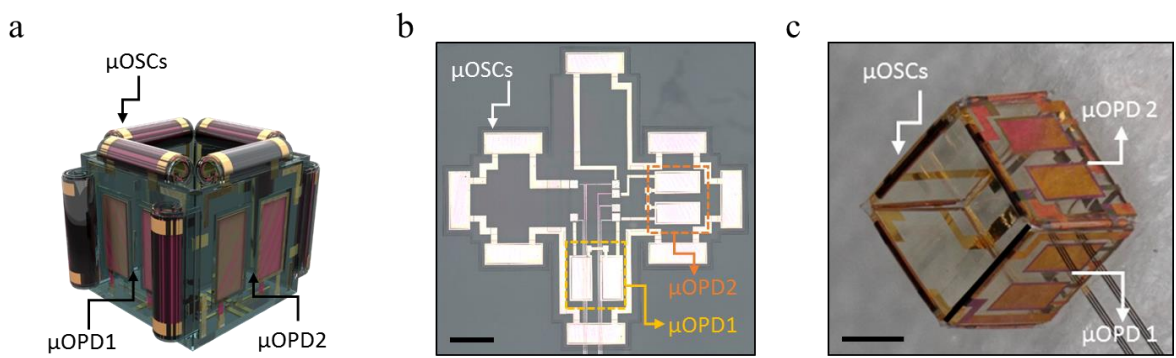

**SI Figure 2 | Smartlet with µOSCs and µOPDs. a,** Schematic illustrating eight integrated µOSCs and two sets of micro organic photodetectors (µOPD1 and µOPD2). Each set of µOPDs is connected in series. **b-c,** Experimental integration of eight µOSCs, µOPD1 and µOPD2 **(b)** before and **(c)** after rolling and folding into the 3D structure. Scale bar, 0.5 mm **(b-c)**.

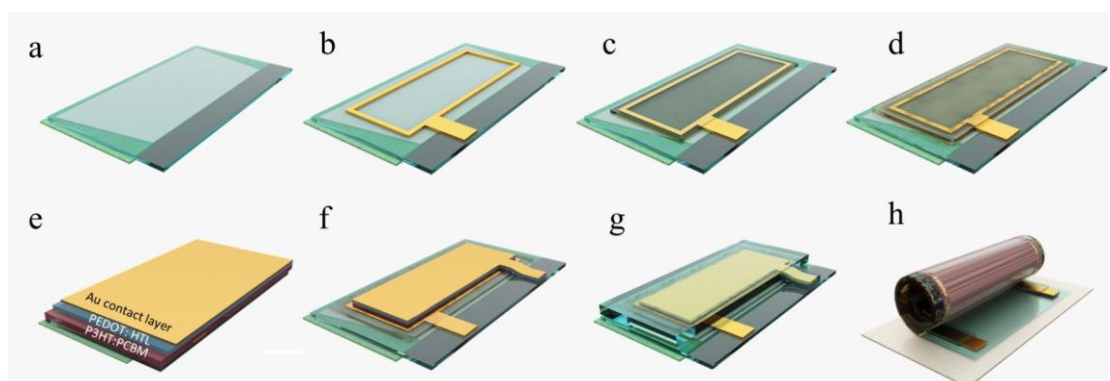

**SI Figure 3 | Schematic illustration of tubular micro organic solar cell (µOSC) showing consecutive fabrication steps. a,** Strained polymeric rolling stack[51,52]. **b,** Bottom electrode contact composed of Cr and Au. **c,** Addition of an ITO transparent anode. **d,** ZnO electron transport layer. **e,** Visualization of top layer stacks comprising a photoactive absorber layer (P3HT:PC$_{61}$BM), hole transport layer (PEDOT:PSS) and Au contact layer. **f,** Photoactive absorber layer (P3HT:PC$_{61}$BM), hole transport layer (PEDOT:PSS) and contact layer (Au) deposited and patterned on the ITO layer. **g,** Passivation layer using SU8 2000.5 photoresist. **h,** Completion of the fabrication process by the self-assembly of the layer stack into a ´Swiss-roll´ structure.

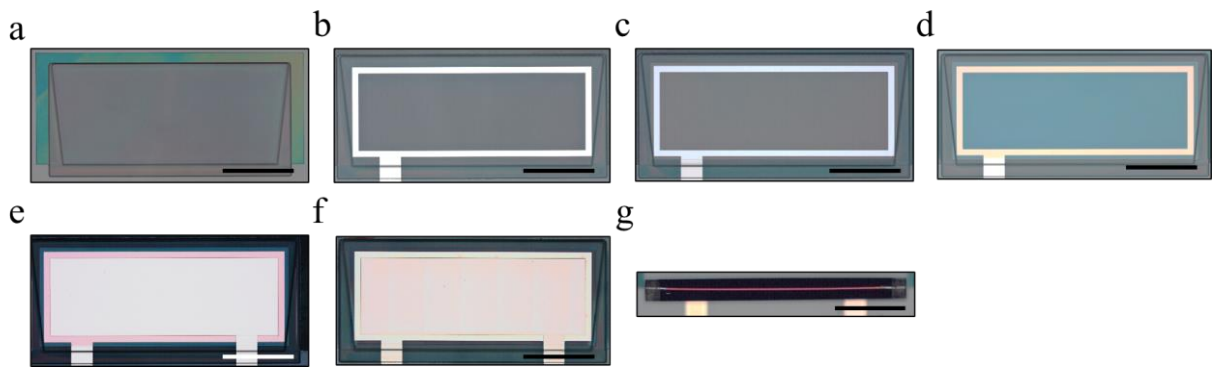

**SI Figure 4 | Microscope images of the consecutive fabrication steps of a tubular μOSC. a,** Strained polymeric rolling stack[59,60]. **b,** Bottom contact (Cr/Au). **c,** Transparent anode (ITO). **d,** Electron transport layer (ZnO). **e,** Photoactive absorber layer (P3HT:PC$_{61}$BM), hole transport layer (PEDOT:PSS) and contact layer (Au). **f,** Passivation layer (SU8 2000.5 photoresist). **g,** Completion of the μOSC fabrication process by forming the Swiss-roll ´structure. Scale bar, 300 μm **(a-g)**.

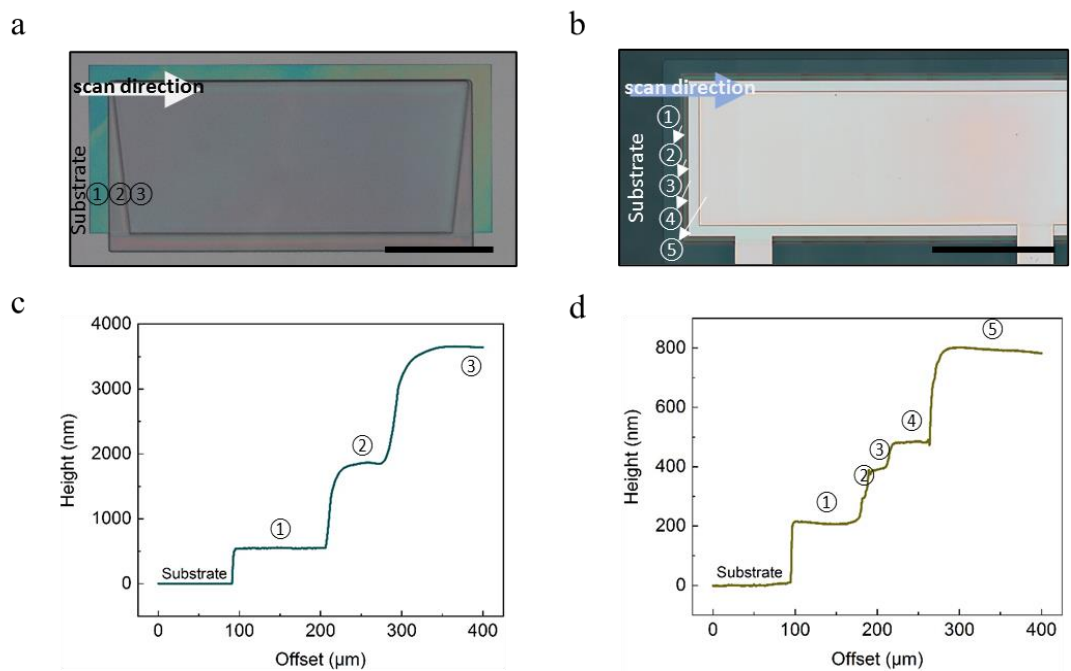

**SI Figure 5 | Thickness analysis of constituent layers for a μOSC. a,** Microscope image of strained polymeric rolling stack. **b,** Microscope image of working layer stack of the μOSC. Scale bars, 300 μm **(a-b)**. **c,** Thickness of strained polymeric rolling stack; ① sacrificial layer, ② polyimide and SU8 passivation layer, and ③ hydrogel layer. **d,** Thickness of working layer stack; ① SU8 passivation layer, ② electron transport layer (ZnO), ③ bottom electrode layer (ITO), ④ contact layer (Cr/Au) and ⑤ photoactive absorber layer (P3HT:PC$_{61}$BM), hole transport layer (PEDOT:PSS) and top contact layer (Au).

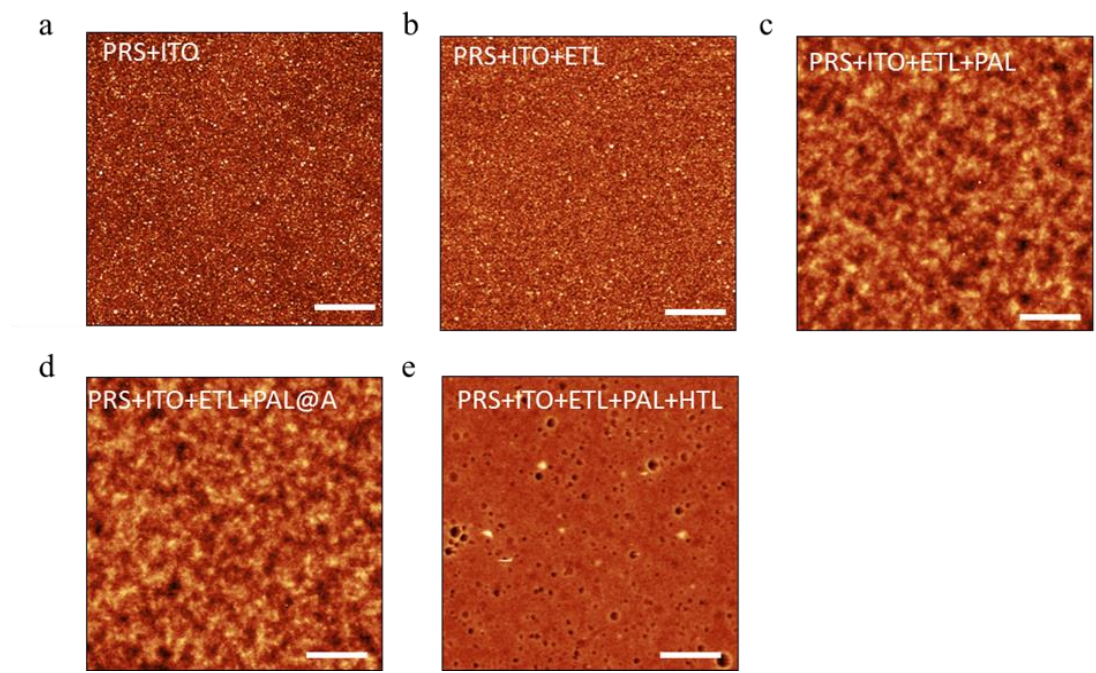

**SI Figure 6 | Atomic force microscopy (AFM) analysis of key constituent layers for a μOSC. a,** 80 nm-thick ITO layer on the polymeric rolling stack (PRS). **b,** 50 nm-thick electron transport layer (ETL, ZnO) on PRS and ITO layer. **c-d,** 200 nm-thick photoactive layer (PAL, P3HT:PC$_{61}$BM) on PRS, ITO and ETL **(c)** before and **(d)** after annealing at 140°C for 10 mins. **e,** 50 nm-thick hole transport layer (HTL, PEDOT: PSS) on PRS, ITO, ETL and PAL. Scale bar, 2 μm **(a-e)**.

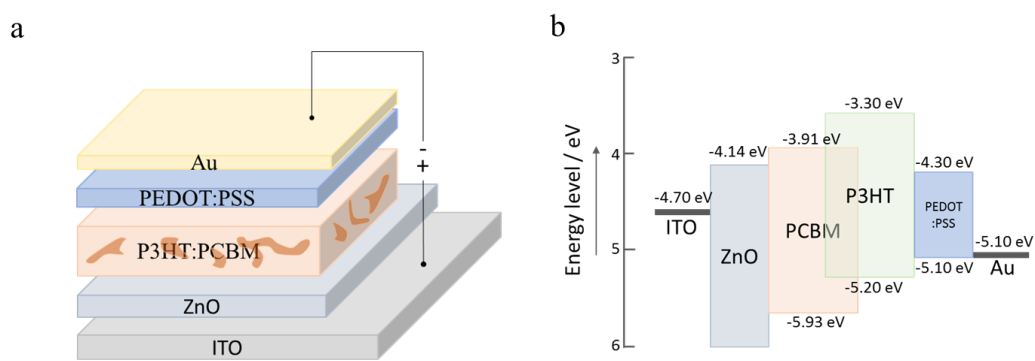

**SI Figure 7 | Schematic illustration of device structure and energy levels. a,** Schematic structure of μOSC based on the P3HT:PC$_{61}$BM photoactive material. **b,** Energy level diagram illustrating the constituents under investigation within the device structure[53-56].

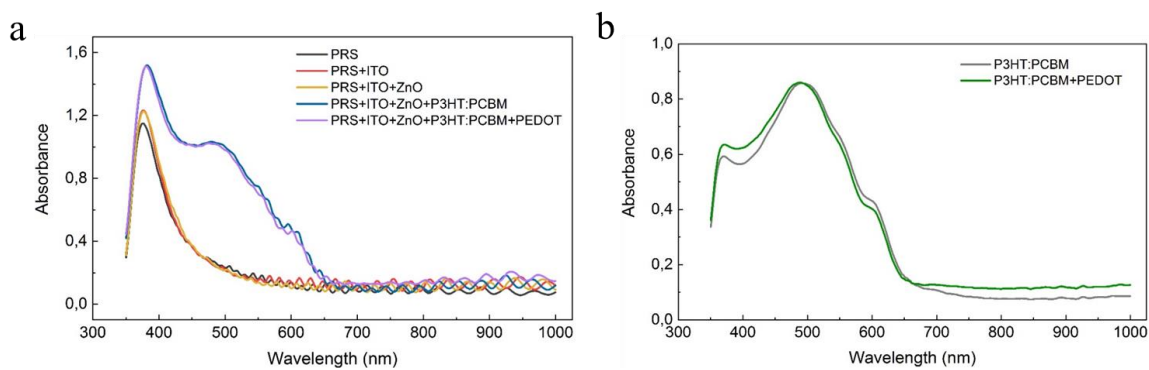

**SI Figure 8 | Absorption spectra of μOSC constituent layers. a,** Analysis of layer-by-layer combinations of μOSC constituent layers: strained polymeric rolling stack (PRS) followed by ITO, ZnO, P3HT:PCBM and PEDOT. Thin film interference causes oscillations in the absorption spectra due to overlapping light wave reflections at different interfaces within the multilayer structure. **b,** Absorption spectra of the P3HT:PC$_{61}$BM and PEDOT:PSS layers on a glass substrate.

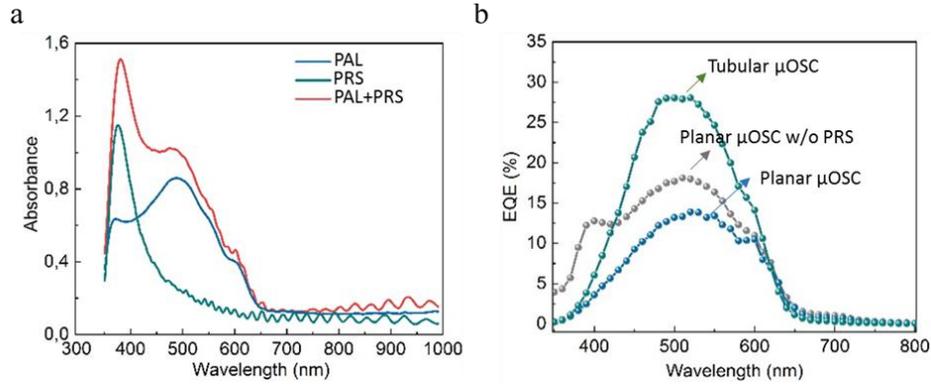

**SI Figure 9 | Absorption and external quantum efficiency (EQE) spectra of μOSC constituent layers. a,** Absorption spectra of photoactive layer (PAL, P3HT:$PC_{61}BM$), polymeric rolling stack (PRS), and combination of PAL and PRS. **b,** EQE spectra of planar μOSC without and with PRS, and tubular μOSC.

### SI Note 1 Current-Voltage (*I-V*) curve characterization

The photovoltaic performance of μOSCs was assessed by the *I-V* curve, utilizing a source meter (KEITHLEY 2636A9). Key parameters such as short-circuit current ($I_{sc}$), open-circuit voltage ($V_{oc}$), maximum power ($P_{max}$), fill factor (FF) and power conversion efficiency (PCE) were derived from the *I-V* curves.

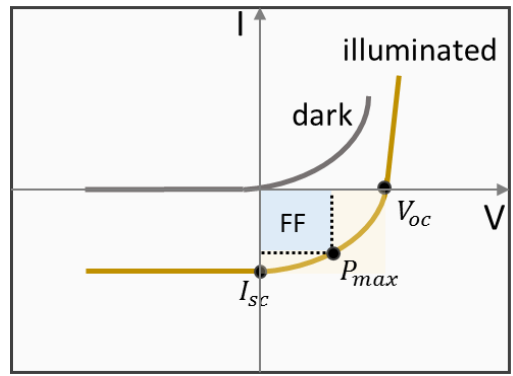

The PCE of the device was determined according to Equation S1:

$$PCE(\%) = \frac{P_{max}}{P_{in} \times A} \times 100 = FF \times \frac{I_{sc}V_{oc}}{P_{in} \times A} \times 100 \quad (1)$$

where the $P_{in}$ represents the incident power density (100 mW $cm^{-2}$) and *A* corresponds to the respective footprint areas of the μOSCs.

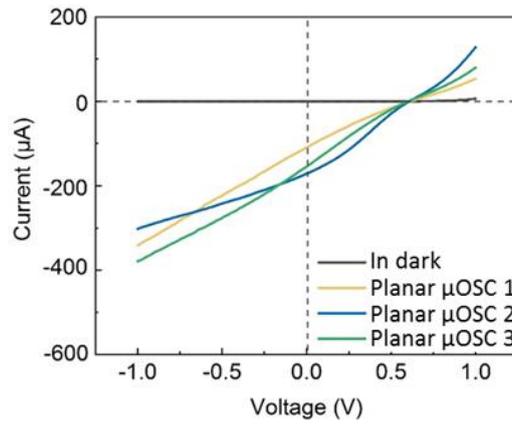

**SI Figure 10 | *I-V* characteristics of planar μOSCs. a,** I-V curves of three different planar μOSCs (All measurements conducted with an illumination of 100 mW cm$^{-2}$).

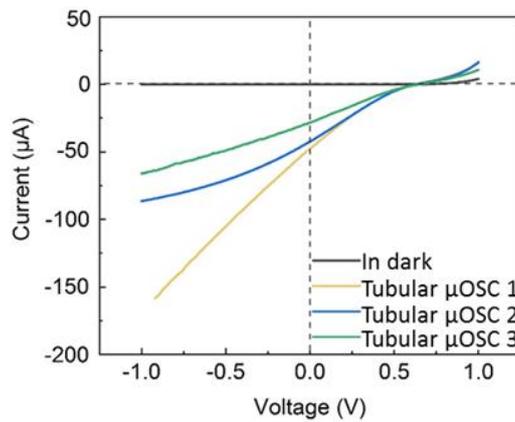

**SI Figure 11 | *I-V* characteristics of tubular μOSCs. a,** I-V curves of three different tubular μOSCs (All measurements conducted with an illumination of 100 mW cm$^{-2}$).

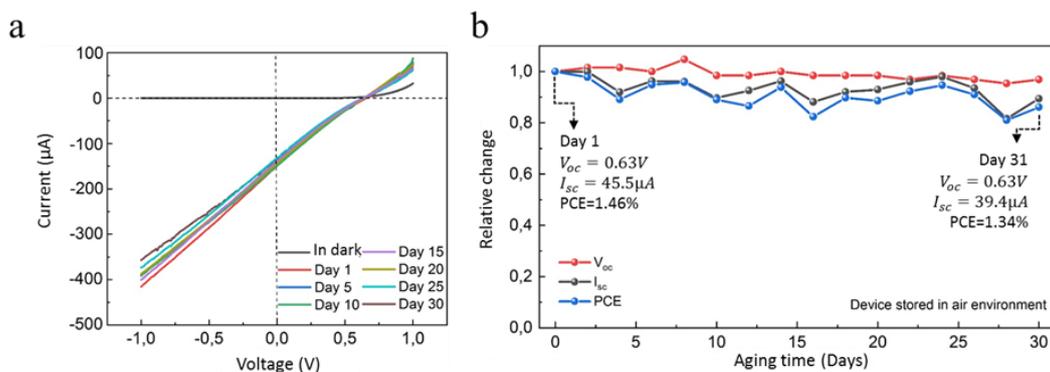

**SI Figure 12 | Lifetime measurement of a planar μOSC over 31 days. a,** *I-V* characteristics measured at intervals of 5 days. **b,** Relative change in $V_{oc}$, $I_{sc}$ and PCE in response to repeated light on and off cycles over 31 days (All measurements conducted with an illumination of 100 mW cm$^{-2}$).

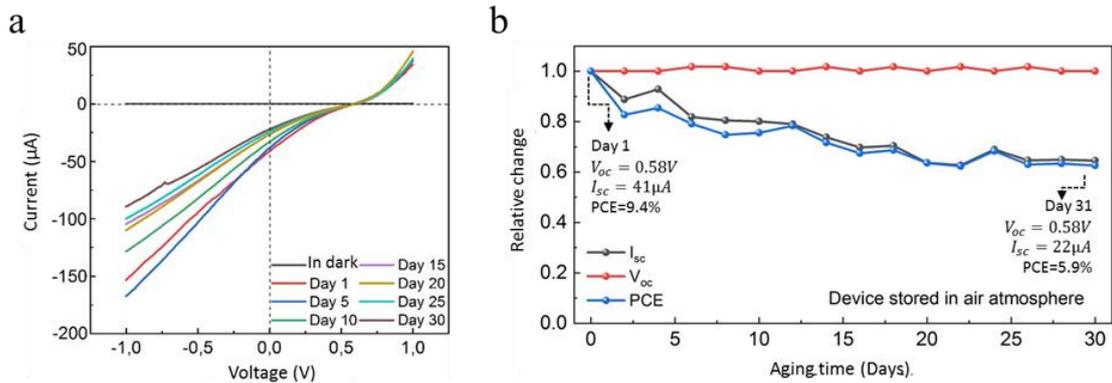

**SI Figure 13 | Lifetime measurement of a tubular µOSC over 31 days. a,** *I-V* characteristics measured at intervals of 5. **b,** Relative change in $V_{oc}$, $I_{sc}$ and PCE in response to repeated light on and off cycles over 31 days (All measurements conducted with an illumination of 100 mW cm$^{-2}$).

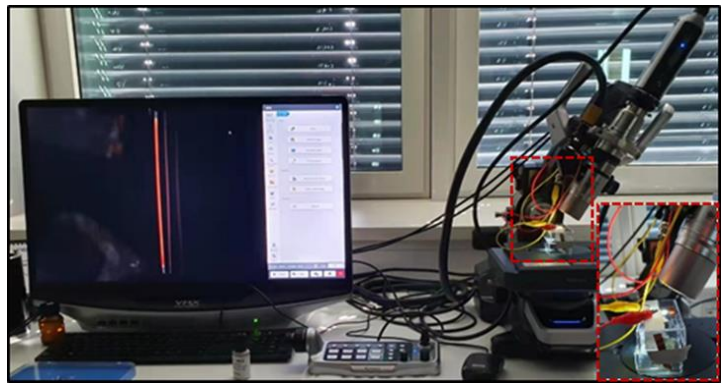

**SI Figure 14 | Angle-dependent measurement setup.** Photo of the measurement setup for angle-dependent analysis. The setup uses a KEYENCE VHX-7000 microscope, which has a custom-integrated optical fiber from a sun simulator. The microscope can be adjusted from 0 ° to 180 °. In the image, the microscope is positioned at 120 °. An inset image in the bottom right corner shows a zoom-in of the sample setup. The light intensity is adjusted and calibrated by modulating the intensity of light source while maintaining the focus on the device to achieve an intensity of 100 mW cm$^{-2}$, measured with an optical light meter (ILT2400, from International Light Technologies)

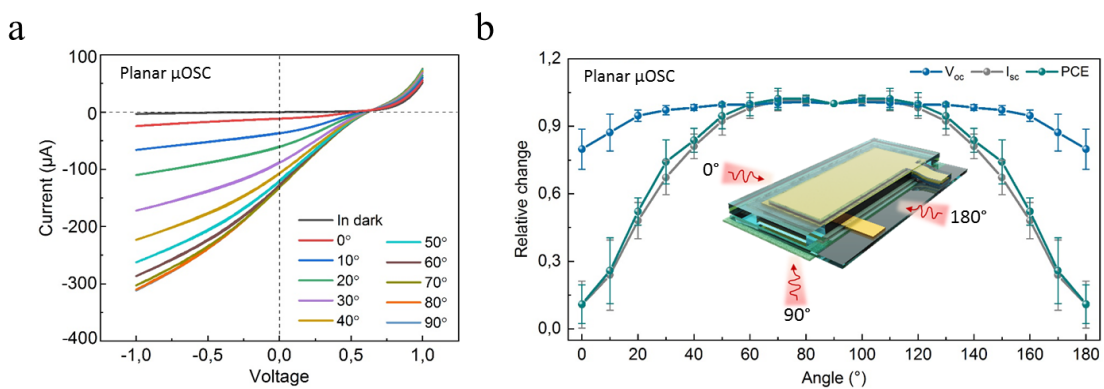

**SI Figure 15 | Effect of tilt angle on the performance of a planar µOSC. a,** *I-V* curves illustrating the performance of a planar µOSC at varying incident angles from 0 ° to 90 °. **b,** Relative changes in $V_{oc}$, $I_{sc}$ and PCE under inclined angles and schematic illustration (inset image) showing the position of incident angles towards the device. Error bars represent the variation in data over three measured devices (All measurements conducted with an illumination of 100 mW cm$^{-2}$).

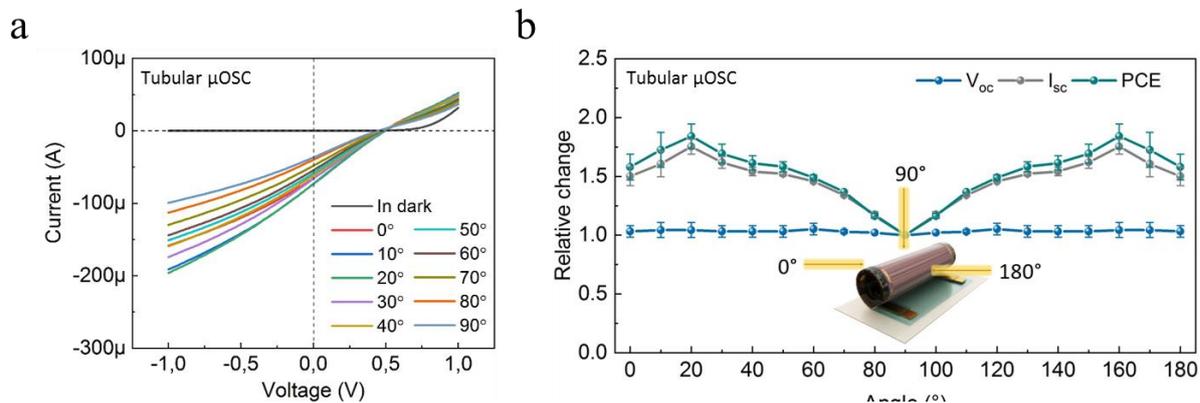

**SI Figure 16 | Effect of tilt angle on the performance of a tubular μOSC. a,** *I-V* curves illustrating the performance of a tubular μOSC at varying incident angles from 0 ° to 90 °. **b,** Relative changes in $V_{oc}$, $I_{sc}$ and PCE under inclined angles and schematic illustration (inset image) showing the position of incident angles towards the device. Error bars represent the variation in data over three measured devices. (All measurements conducted with an illumination of 100 mW cm$^{-2}$).

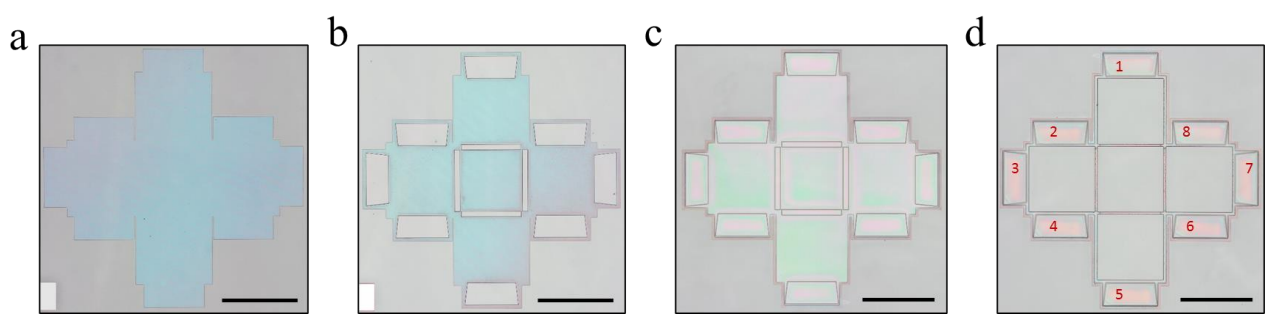

**SI Figure 17 | Microscope images of consecutive fabrication steps towards a polymeric platform with eight μOSCs. a,** Sacrificial layer. **b,** Hydrogel hinge and swelling layers for subsequent folding and rolling processes, respectively. **c,** Supporting polyimide layer. **d,** Stiffening SU8 photoresist layer. Scale bar, 1 mm **(a-d)**.

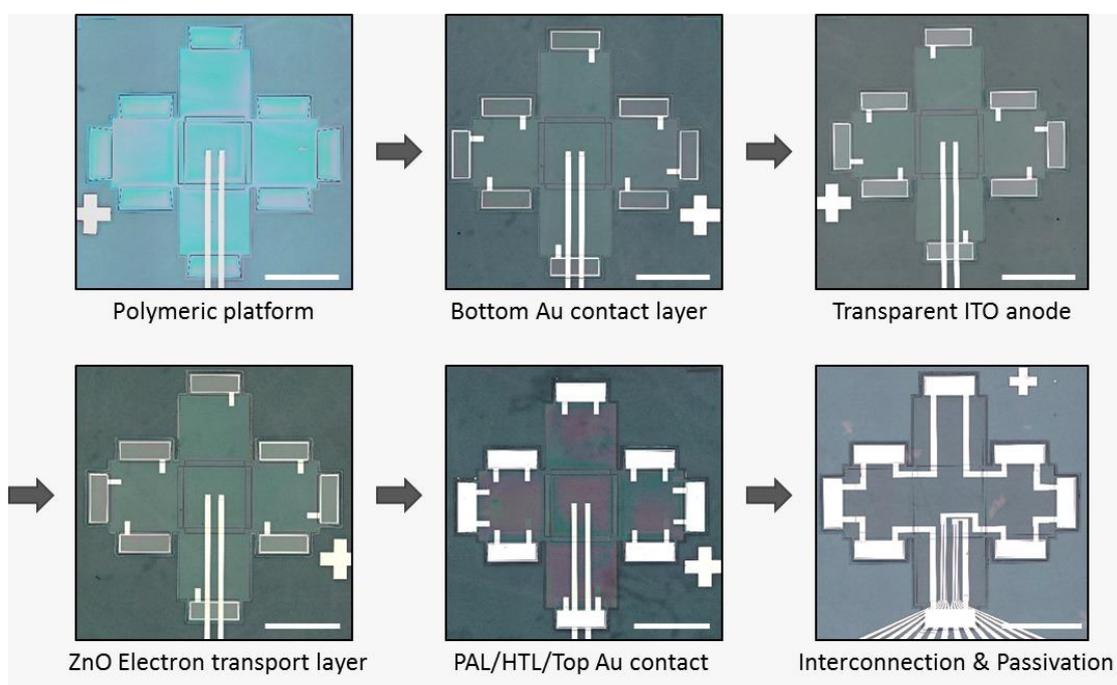

**SI Figure 18 | Microscope images of consecutive fabrication steps towards a planar smartlet with eight micro-organic solar cells (μOSCs).** Fabrication process of μOSCs on the polymeric platform, followed by bottom Au contact layer deposition for interconnections, transparent ITO anode, ZnO deposition as an electron transport layer (ETL), photoactive layer (PAL) with a hole transport layer (HTL) and top Au contact layer deposition, and finally deposition of Au interconnection and passivation by SU-8 photoresist. Scale bar, 1 mm.

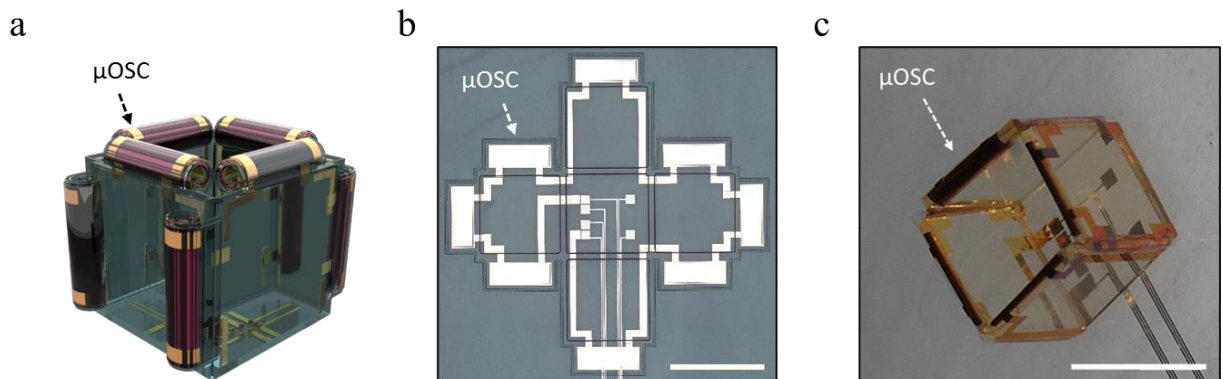

**SI Figure 19 | Smartlet with tubular µOSCs serially connected and located at eight edges. a,** Schematic illustrating eight integrated tubular µOSCs on the smartlet edges. **b-c,** Experimental integration of eight µOSCs connected in series **(b)** before and **(c)** after rolling and folding into a smartlet. Scale bar, 1 mm **(b-c)**.

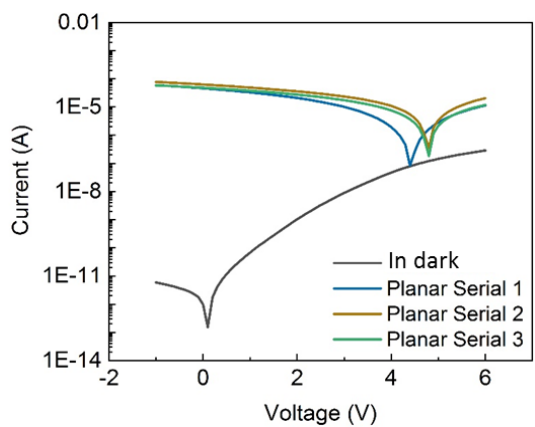

**SI Figure 20 | Photovoltaic I-V curves for serially connected µOSCs. a,** I-V curves of three different planar µOSCs serially connected (All measurements conducted with an illumination of 100 mW cm$^{-2}$).

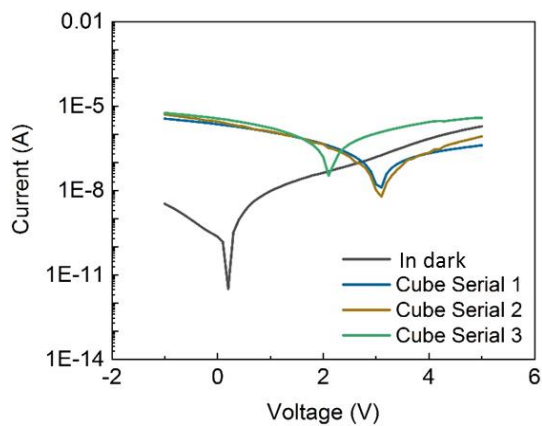

**SI Figure 21 | Photovoltaic characteristics I-V curves for serially connected µOSCs. a,** I-V curves of three different rolled and folded µOSCs serially connected (All measurements conducted with an illumination of 100 mW cm$^{-2}$).

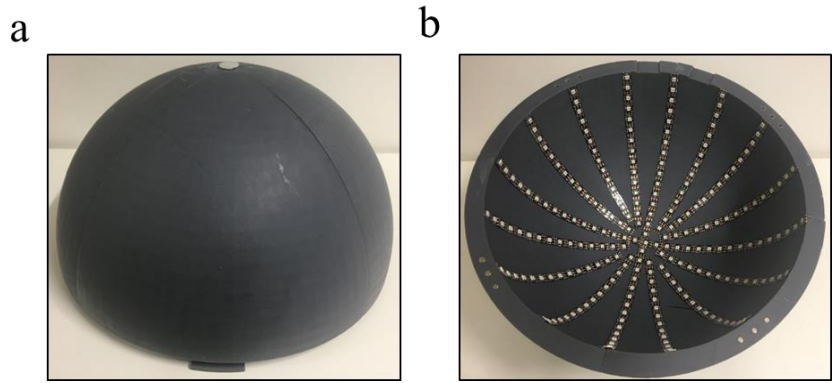

**SI Figure 22 | 3-axis angle dependent measurement setup for μOSCs in the smartlet. a-b,** Photographs of the **(a)** outside and **(b)** inside of the 3D light dome with 240 commercial LEDs placed at every azimuth (altitude) angle in steps of 22.5° (17.5°).

### SI Note 2 Project source code

The LEDs within the 3D light dome are controlled through a printed circuit board that includes an Arduino UNO Rev3 microcontroller. The Arduino code enables serial communication to receive commands for toggling individual LEDs. Each LED is addressed via a cascade LED driver on the NeoPixel LED strip, which is controlled by the Arduino. The code toggles each LED's state (ON/OFF) based on the received commands, allowing for precise control over the illumination pattern. The LEDs are addressed by their index, which corresponds to their physical position within the 3D light dome.

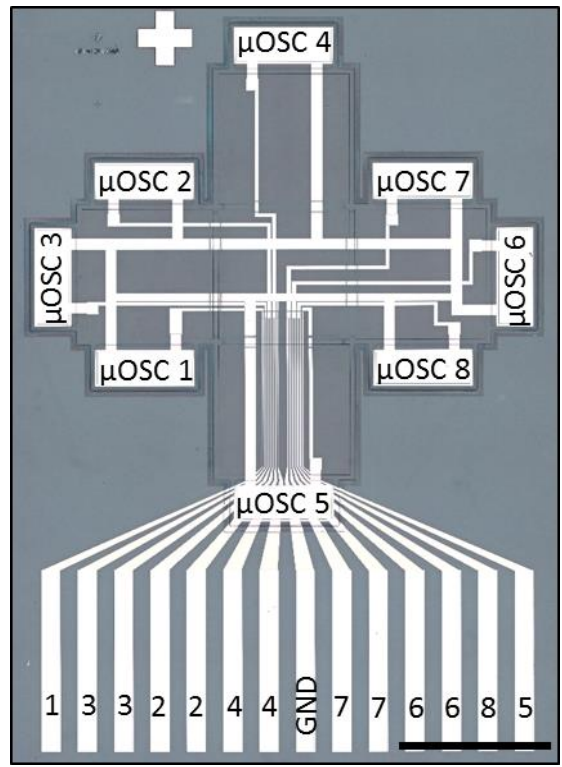

**SI Figure 23 | Microscope image of μOSCs on the planar smartlet for incident angle and position dependent measurements.** A total of eight μOSCs are individually connected to measurement pads. Scale bar, 1 mm.

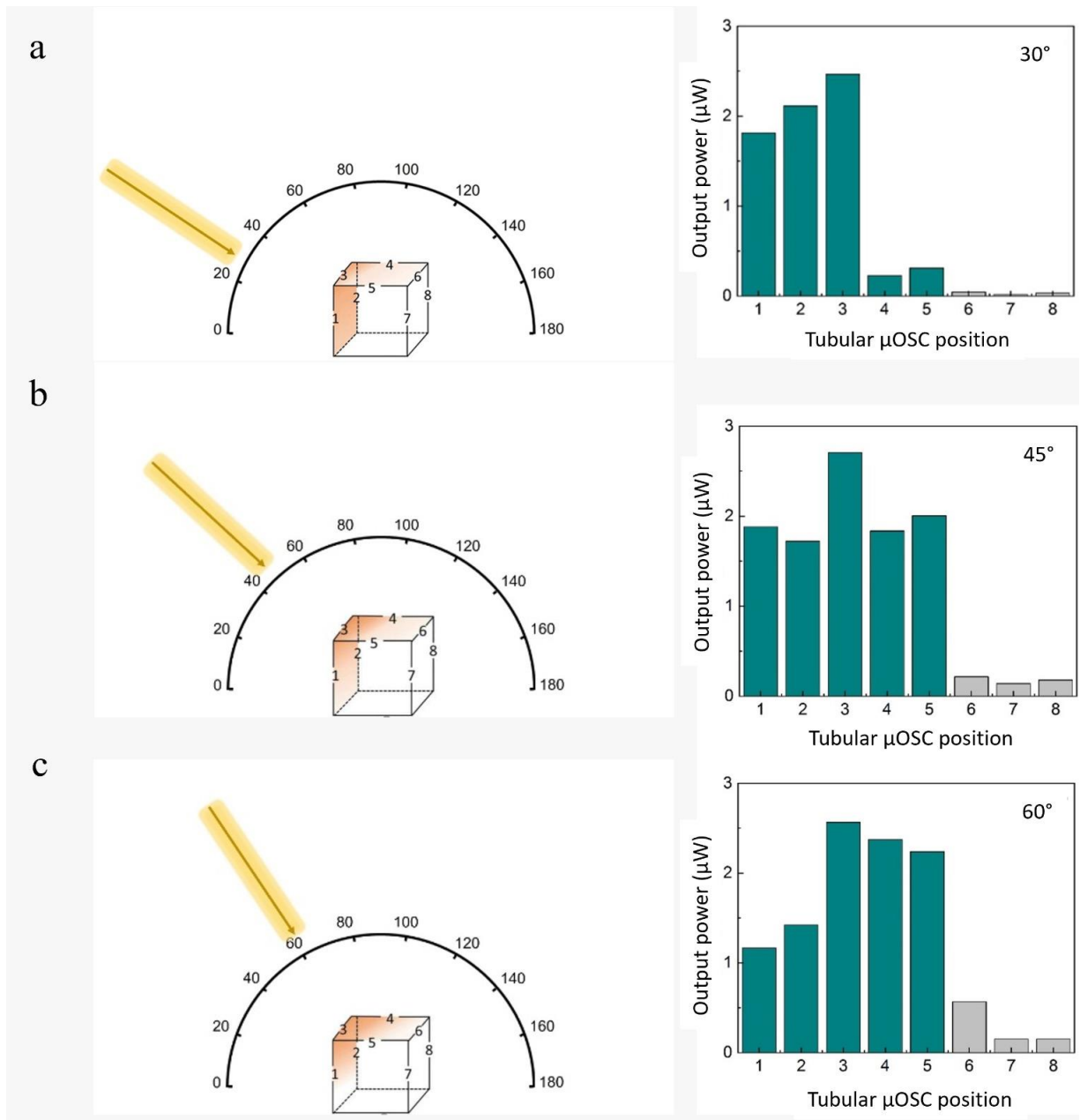

**SI Figure 24 | Incident angle and position dependent measurements of µOSCs. a-c,** Schematic illustration of incident angles and eight smartlet integrated tubular µOSCs (left) and output power measured at incident angles of 30°, 45° and 60°(right) (All measurements with the 3D light dome conducted with an illumination of 50 µW cm$^{-2}$).

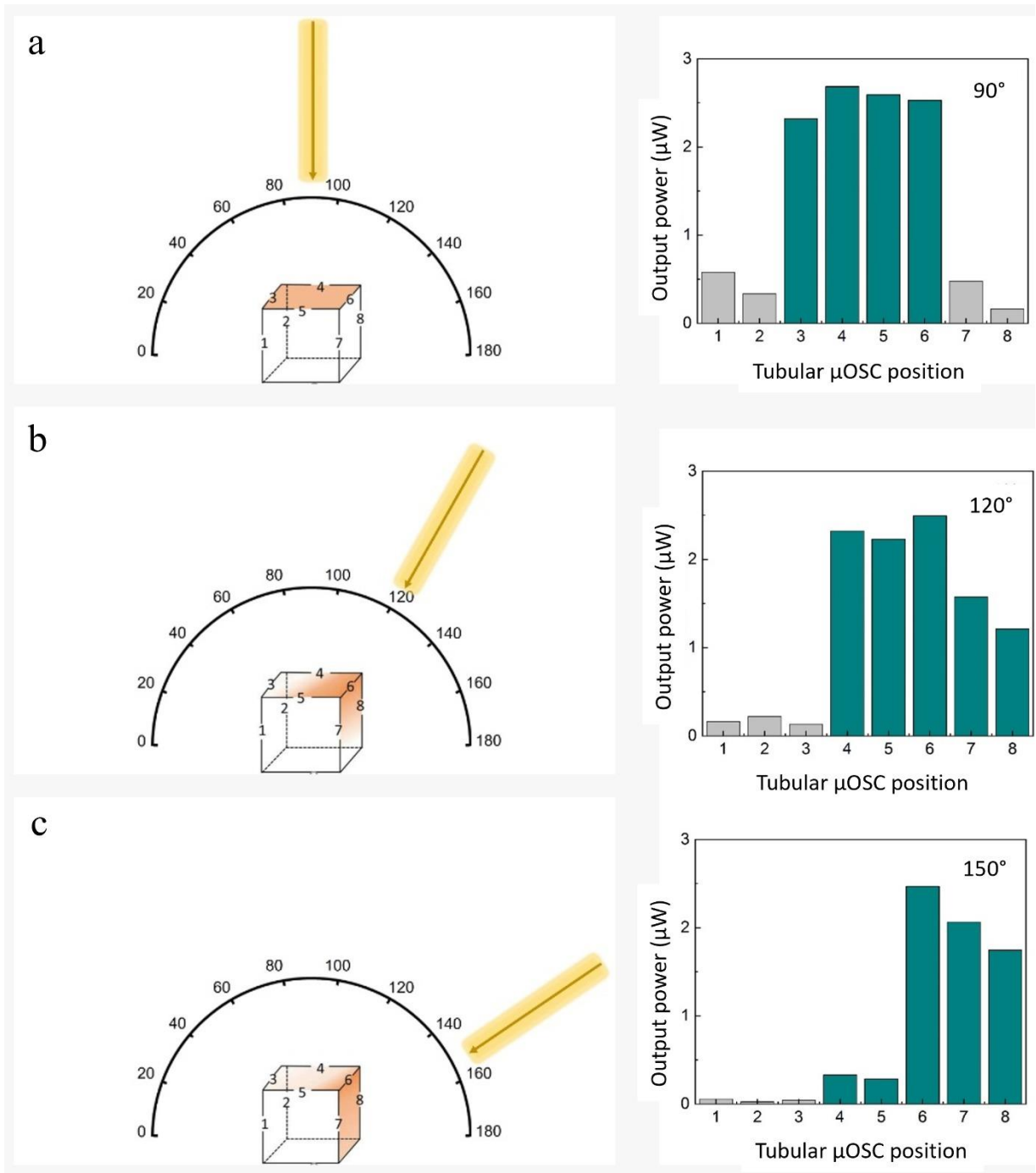

**SI Figure 25 | Incident angle and position dependent measurements of μOSCs. a-c,** Schematic illustration of incident angles and eight smartlet integrated tubular μOSCs (left) and output power measured at incident angles of 90°, 120° and 150° (right) (All measurements with the 3D light dome conducted with an illumination of 50 μW cm$^{-2}$).

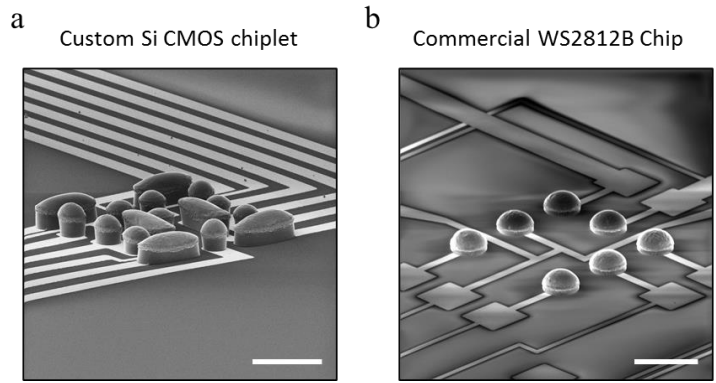

**SI Figure 26 | Optimized Cu/Sn pillar bumps for subsequent μChip solid-liquid interdiffusion (SLID) bonding based on contact layout of μChips.** SEM images of two distinct Cu/Sn pillar bump layouts for **(a)** customized silicon CMOS chiplet and **(b)** commercial WS2812B chip. Scale bar, 100 μm and 50 μm, respectively.

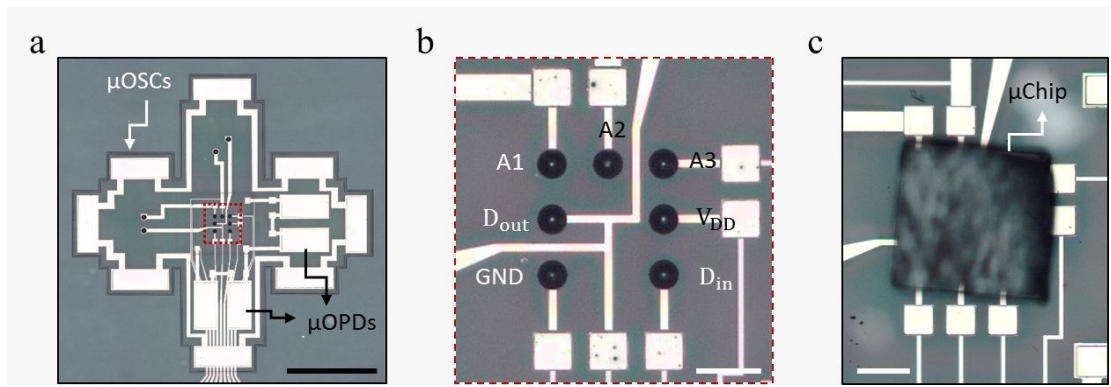

**SI Figure 27 | Microscope images of planar smartlet including μOSCs, μOPDs and μChip (commercial WS2812B chip) on Cu/Sn pillar bump layout. a,** Integrated μOSCs, μOPDs and Cu/Sn pillar bumps for subsequent bonding of μChip. Scale bar, 1 mm. **b,** Magnified view of Cu/Sn bump array illustrating the interconnection of the pads: Ground (GND), $V_{DD}$ (power supply), $D_{out}$ (control data signal output), $D_{in}$ (control data signal input) and three actuation pads (A1, A2, A3). Scale bar, 200 μm. **c,** Integrated μChip after bonding onto the Cu/Sn pillar array. Scale bar, 100 μm.

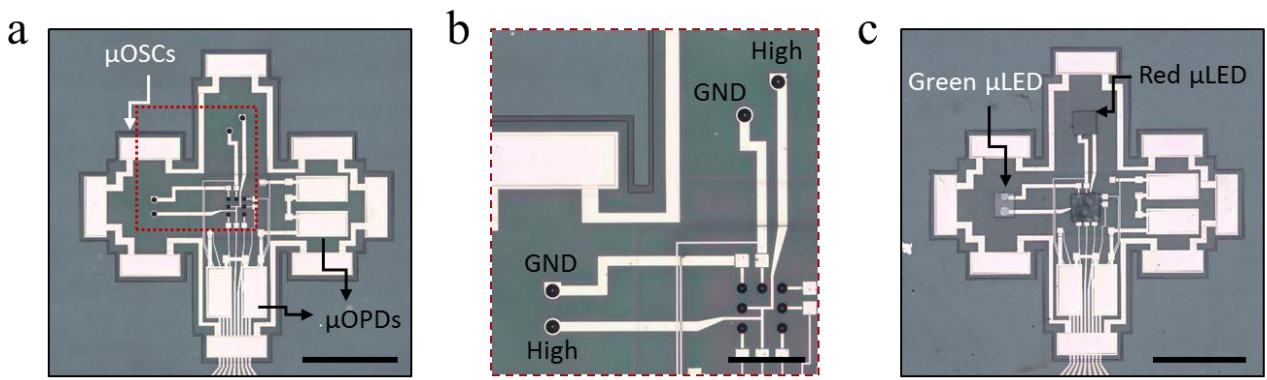

**SI Figure 28 | Microscope images of planar smartlet including μOSCs, μOPDs, μChip and μLEDs. a,** Planar smartlet with μOSCs, two μOPDs and Cu/Sn pillar bump layout for bonding of μChip and μLED. Scale bar, 1 mm. **b,** Magnified view of Cu/Sn bump layout and interconnection lines. Scale bar, 400 μm. **c,** Integration of μChip (commercial WS2812B chip), green and red μLEDs after bonding on Cu/Sn pillar bumps. Scale bar, 1 mm.

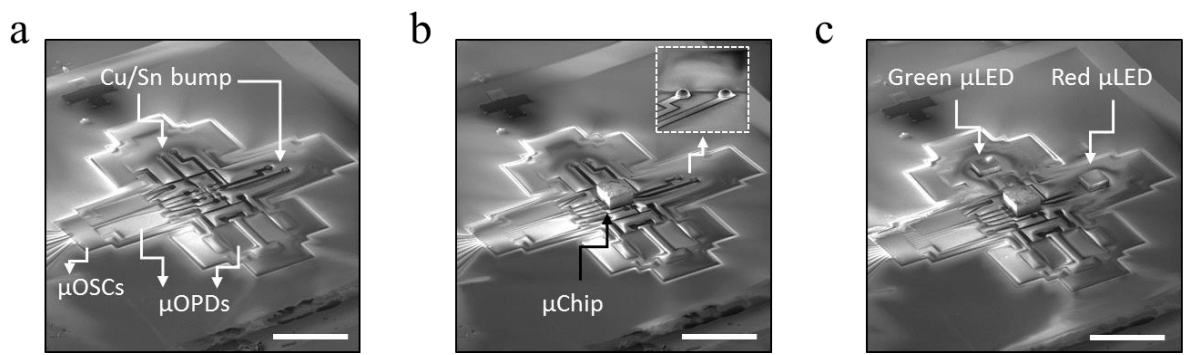

**SI Figure 29 | SEM images of planar smartlet integrating μOSCs, μOPDs, μChip and μLEDs. a,** Planar smartlet with μOSCs, μOPDs and Cu/Sn pillar bumps highlighting those for μLED bonding. **b,** Intermediate stage of planar smartlet after bonding of the μChip onto the Cu/Sn pillar bump array in the middle and zoomed inset view of Cu/Sn pillar bumps for red μLED bonding. **c,** Complete integration of μOSCs, μOPDs, μChip (commercial WS2812B chip) and green and red μLEDs following bonding on the Cu/Sn pillar bumps. Scale bar, 1 mm (a-c).

## SI Note 3 Description of Lablet μChips

Lablets were designed by the groups of J.S.M. and J. Öhm, and fabrication coordinated by J.S.M and P. Mayr all at Ruhr University Bochum as part of the EU MICREAgents Project (2012-2016)[56,57]. Briefly, Europractice engineering runs of 200mm Si wafers on the 180nm TSMC node yielded wafers with MLM (4 offset masks in one) reticules of size 11.2x14.9 mm. The Al pads on the wafers were plated in a custom electroless procedure (Atotech GmbH, Berlin) with a Cr, Ni, Pd, Au stack (75nm gold) to fill the 1.2 μm insulation cavities of the pads to approximately planarize the surface of the lablets. The reticules consisted of 4 main chips: a docking chip consisting of an 128x128 array of 2x2 electrode pixels with smart sensor and actuator electrodes, and three arrays of lablets (each with ca. 1000 lablets). The first and individually addressable lablet array was used to test individual lablets as part of the array, using a 15-signal bus. The other two lablet arrays were connected globally by a 5-signal bus, signals separately supplying power and connecting 2 sensors and 2 supercapacitor pads for global galvanic processing. These two chip types were tested as a whole and later used at TU Chemnitz, MAIN as a source from which to singulate individual thinned lablets. The severed buried lateral bus connections were etched peripherally using aqua regis to eliminate any trace of lateral metal. Final lablets in the 3rd project run (CMOS3) had dimensions of 140x140x35 μm and supported 14 pads (including 6 electrodes for power: 2 for harvesting and 4 for two supercapacitors (requiring galvanic post-processing). For MICREAgents, the wafers were extensively post-processed[64,65] as autonomous electrochemical lablets. In this work, we have utilized the bare electroless plated lablets only. Singulation and thinning at chip scale for this work were performed by pulsed laser dicing (15 μm grooves, ca. 48 μm deep, TOPAG Lasertechnik GmbH) and back polishing to 50 μm (Fraunhofer ENAS).

The ca. 1000 lablets in each of the arrays were designed combinatorially, as variants with two different clock speeds (nominally 20 Hz and 400 Hz), five different pad geometries, and 5 variant programmable finite state machines which differed in several respects including (i) autorun: the program started autonomously after a timeout or only when externally triggered (ii) details of the Manchester encoding communication including pulse lengths and whether biased or bipolar (iii) the default initial program loaded onto the chip. The 58-bit program for the lablets is described in[65], and implements a four-phase processing controlling various repeats of 8-cycle patterns of tristate or active signals on the up to 6 actuator electrodes, and conditional switching between phases depending on digital signals derived from the two electrochemical sensors. Other features of the programmable finite state machine, including the ability to record sensor data and to replicate itself between lablets are not used in the current work.

## SI Note 4 Programming of lablet µChips

The lablets use a Manchester code for asynchronous programming and communication, and different lablet variants are configured via different clock frequencies and Manchester coding parameters to respond at different programming frequencies, which will prove useful in the context of smartlet collectives below. The programmable state machine of the lablet involves an idle state and three active phases, with a programmable number of repeats and clock rate in each phase for its programmable 8-step binary + tristate actuator patterns. Transitions between four phases are cyclic in the absence of programmable conditions. These are separately programmable in each phase, dependent on two sensor inputs and/or the receipt of a trigger command, and result in a programmed transition to the previous, same, next or idle phases. The end of each can also be programmed to send a command or to replicate their program with accumulated data to the data output. In this work, this output is connected to a green µLED on the smartlet, to send data, commands or the program to another smartlet (or to the operator). The characterization of lablets is provided in **SI Fig. 30**.

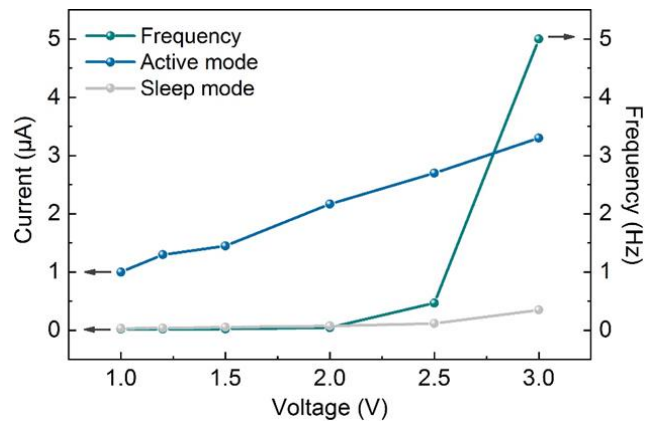

**SI Figure 30 | Characterization of bonded customized Si CMOS chiplet on Cu/Sn pillar bumps.** Operation frequency and output current of the Si-Chiplet 1 as a function of applied $V_{DD}$ in sleep mode and activation mode.

## SI Note 5 Description of the commercial LED driver chiplet

The commercial LED driver chiplet (WS2812B, datasheet from Worldsemi Co.) has the following electrical and switching characteristics within a temperature window of
$T_A$ = -20~+70°C, power supply $V_{DD}$=4.5~5.5V and ground $V_{SS}$=0V.

Electrical characteristics:

| Parameter | Symbol | conditions | Min | Typ | Max | Unit |
|---|---|---|---|---|---|---|
| Input current | $I_I$ | $V_I$=$V_{DD}$/$V_{SS}$ | | | $\pm 1$ | µA |
| Input voltage level | $V_{IH}$ | $D_{IN}$, SET | 0.7 $V_{DD}$ | | | V |
| | $V_{IL}$ | $D_{IN}$, SET | | | 0.3 $V_{DD}$ | V |
| Hysteresis voltage | $V_H$ | $D_{IN}$, SET | | 0.35 | | V |

Switching characteristics:

| Parameter | Symbol | conditions | Min | Typ | Max | Unit |
|---|---|---|---|---|---|---|
| Transmission delay time | $t_{PLZ}$ | CL=15pF, $D_{IN}$→$D_{OUT}$, RL=10KΩ | | | 300 | ns |
| Fall time | $t_{THZ}$ | CL=300pF, OUTR/OUTG/OUTB | | | 120 | µs |
| Input capacity | $C_t$ | | | | 15 | pF |

Data transfer time (TH+TL=1.25 μs ± 600 ns):

| | | | |
|---|---|---|---|
| T0H | 0 code, high voltage time | 0.4 μs | ±150 ns |
| T1H | 1 code, high voltage time | 0.8 μs | ±150 ns |
| T0L | 0 code, low voltage time | 0.85 μs | ±150 ns |
| T1L | 1 code, low voltage time | 0.45 μs | ±150 ns |
| RES | Low voltage time | Above 50 μs | |

Sequence chart:

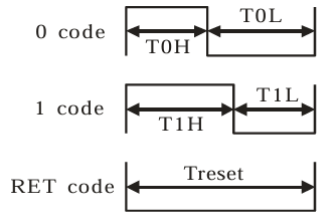

Cascade method:

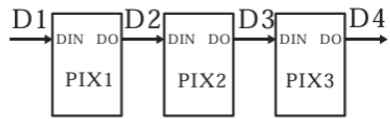

Data transmission method:

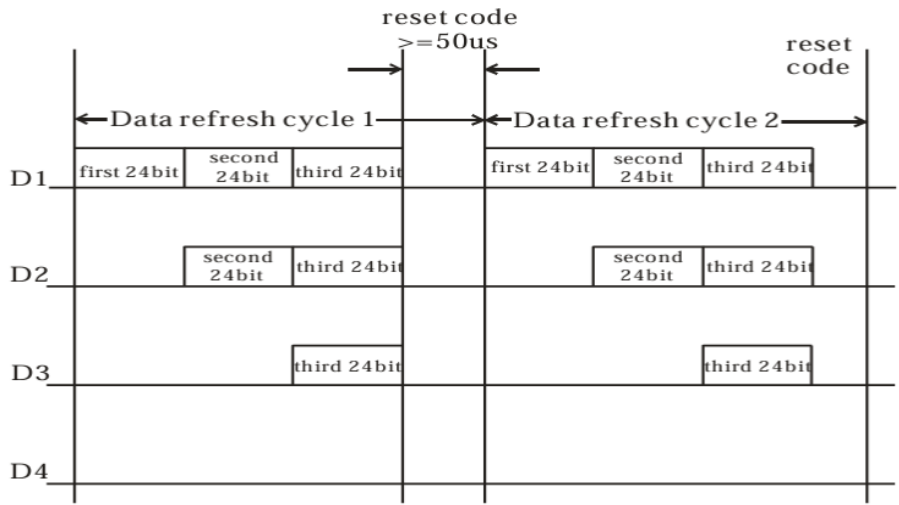

The data of D1 is send by MCU, and D2, D3, D4 through pixel internal reshaping amplification to transmit.

Composition of 24-bit data:

| $A_17$ | $A_16$ | $A_15$ | $A_14$ | $A_13$ | $A_12$ | $A_11$ | $A_10$ | $A_27$ | $A_26$ | $A_25$ | $A_24$ | $A_23$ | $A_22$ | $A_21$ | $A_20$ | $A_37$ | $A_36$ | $A_35$ | $A_34$ | $A_33$ | $A_32$ | $A_31$ | $A_30$ |
|---|---|---|---|---|---|---|---|---|---|---|---|---|---|---|---|---|---|---|---|---|---|---|---|

Follow the order of $A_1$, $A_2$, $A_3$ to send data and the high bit sent at first

**SI Note 6 Programming of LED driver μChips**

The commercial LED driver chiplet (WS2812B, from Worldsemi Co.) is a control circuitry including an internal intelligent digital data latch and signal reshaping amplification drive circuit, a precision internal oscillator, and a 12V voltage programmable constant current control component. The data transfer protocol employs a single non-return-to-zero (NZR) communication mode. Upon power-on reset, the control data signal input (Din) port receives data from the controller. The first pixel collects an initial 24-bit data set and sends it to the internal data latch. Subsequent data, reshaped by the internal signal reshaping amplification circuit, is transmitted to the next pixel via the control data signal output (Dout) port. Each pixel transmission reduces the signal by 24 bits. The auto reshaping transmit technology of the pixels allows for an unlimited number of cascaded pixels, constrained only by the signal transmission speed.

**SI Note 7 Detailed description of μLED chips**

The commercial μLED chips (datasheet from EPIGAP Optronic GmbH, Germany) have the following optical and electrical characteristics at room temperature (25 ℃).

Green μLED (EOLC 515-35):

| Parameter | Test cond. | Min | Typ | Max |
|---|---|---|---|---|
| Forward voltage | 1 μA | 1.7 V | | |
| Forward voltage | 20 mA | 2.8 V | | 3.8 V |
| Reverse current | 8 V | | | 0.5 μA |
| Dominant wavelength | 20 mA | | 515 nm | |
| Luminous intensity | 20 mA | 485 mcd | | 530 mcd |

Chip thickness 90 ± 10 μm, Chip size 330 ± 20 μm × 275 ± 20 μm

Red μLED (EOLC 635-34):

| Parameter | Test cond. | Min | Typ | Max |
|---|---|---|---|---|
| Forward voltage | 20 mA | 1.8 V | | 2.5 V |
| Reverse current | 10 V | | | 5 μA |
| Peak wavelength | 20 mA | | 635 nm | |
| Dominant wavelength | 20 mA | 619 nm | | 629 nm |
| Luminous intensity | 20 mA | 500 mcd | | 750 mcd |

Chip thickness 100 ± 15 μm, Chip size 330 × 300 ± 25 μm

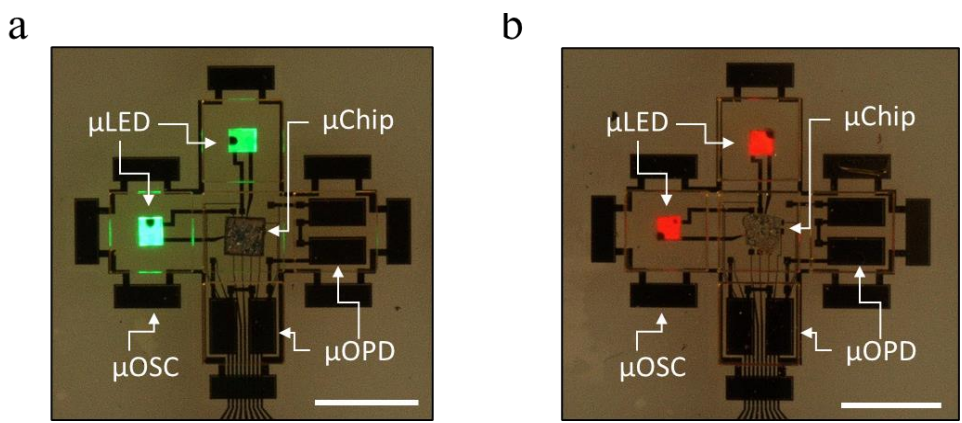

**SI Figure 31 | Microscope images of planar smartlets integrating μOSCs, μOPDs, μChip and μLEDs. a,** Green μLEDs. **b,** Red μLEDs. Scale bar, 1 mm (a-b).

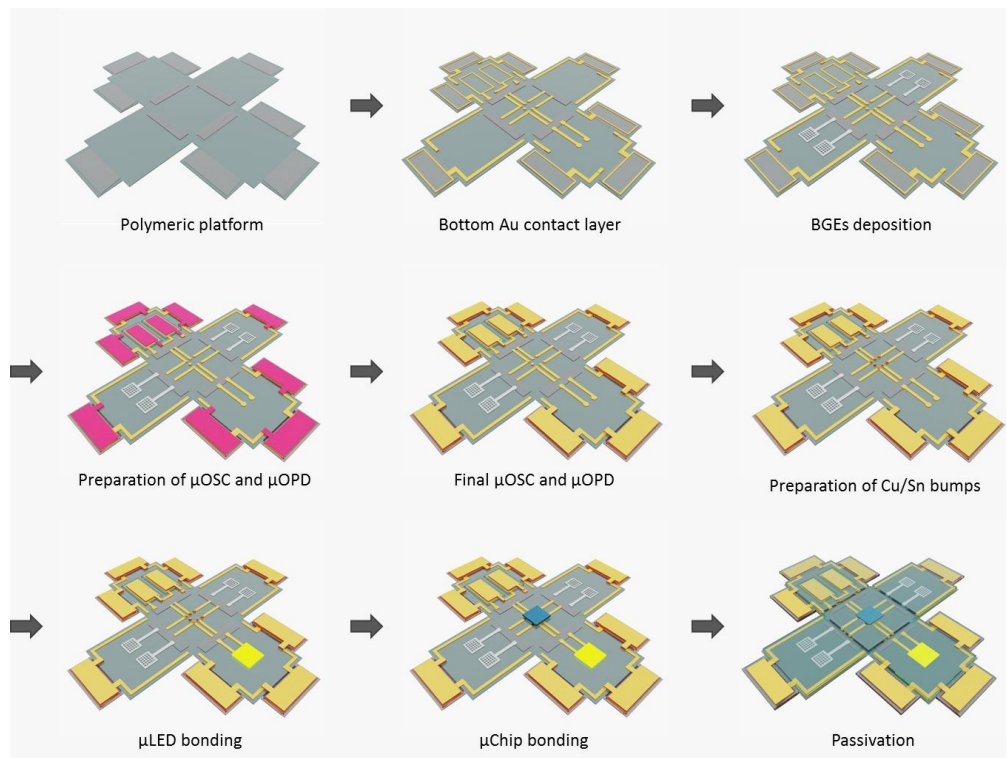

**SI Figure 32 | Schematic illustration of the fabrication steps towards the fully equipped planar smartlet.** The fabrication process begins with the creation of the polymeric platform, followed by bottom Au contact layer deposition for interconnections, deposition of Pt and Ni for bubble generating electrodes (BGEs) functionality, intermediate steps for μOSC and μOPD fabrication including anode, electron transport layer, photoactive polymer and hole transport layer. The final steps involve Au deposition for both μOSCs and μOPDs, preparation of Cu/Sn bumps, bonding of μLEDs and μChips, and finally SU-8 passivation.

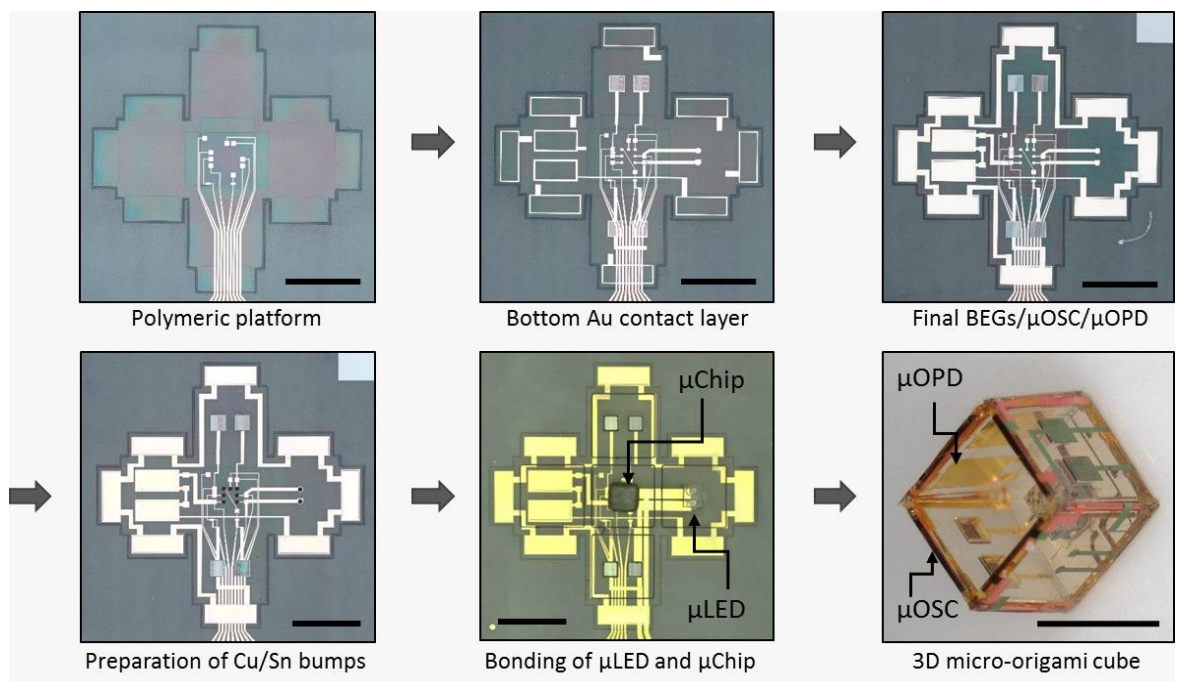

**SI Figure 33 | Microscope images of the fabrication steps towards the fully equipped smartlet.** The fabrication process begins with the creation of a polymeric platform, followed by Au deposition for interconnections, complete fabrication of BGEs, μOSCs and μOPDs, preparation of Cu/Sn bumps, bonding of μLEDs and μChip with SU-8 passivation, and finally 3D self-assembled smartlet with all components integrated. Scale bar, 1 mm.

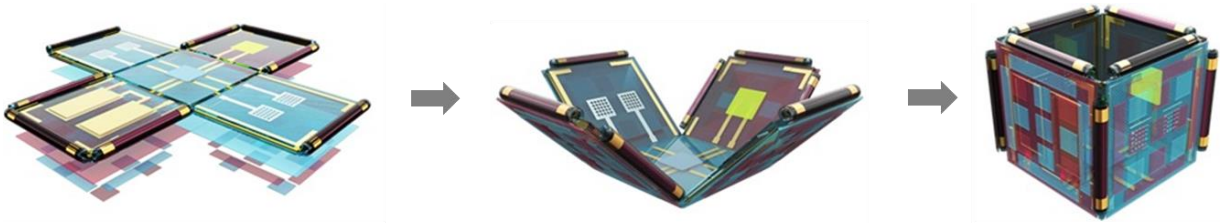

**SI Figure 34 | Fabrication and patterning of hydrophobic-hydropillic layer.** Schematic illustration of the hydrophobic-hydrophilic patterning of the outer surfaces of the smartlet in 2D and self-assembly into 3D smartlet.

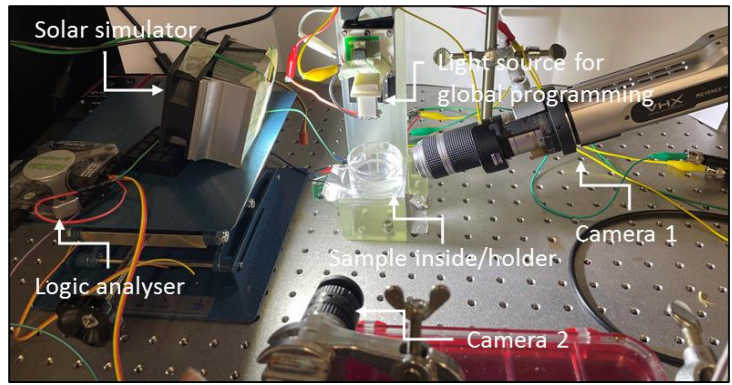

**SI Figure 35 | Measurement setup for locomotion.** Experimental setup including solar simulator, logic analyzer for programming the μChip to regulate actuation, additional light source for global programming by sending a program through light, sample holder for samples within a beaker and two cameras positioned at different angles to capture the locomotion process from multiple perspectives.